\documentclass{emulateapj}
\usepackage{epsfig}
\usepackage{multirow}
\usepackage[fleqn]{amsmath}

\begin{document}

\title{A Spectroscopic and Photometric Exploration of the C/M Ratio in the Disk of M31}
\author{Katherine M. Hamren\altaffilmark{1,2}, Constance M. Rockosi\altaffilmark{1}, Puragra Guhathakurta\altaffilmark{1}, Martha L. Boyer\altaffilmark{3}, Graeme H. Smith\altaffilmark{1}, Julianne J. Dalcanton\altaffilmark{4}, Dylan Gregersen\altaffilmark{5}, Anil C. Seth\altaffilmark{5}, Alexia R. Lewis\altaffilmark{4}, Benjamin F. Williams\altaffilmark{4}, Elisa Toloba\altaffilmark{1}, L\'{e}o Girardi\altaffilmark{6},Claire E. Dorman\altaffilmark{1}, Karoline M. Gilbert\altaffilmark{7}, Daniel R. Weisz\altaffilmark{4,8}}

\altaffiltext{1}{Department of Astronomy and Astrophysics, University of California Santa Cruz, 1156 High Street, Santa Cruz, CA  95064, USA}
\altaffiltext{2}{{\tt khamren@ucolick.org}}
\altaffiltext{3}{Observational Cosmology Lab, Code 665, NASA Goddard Space Flight Center, Greenbelt, MD 20771, USA}
\altaffiltext{4}{Department of Astronomy, Box 351580, University of Washington, Seattle, WA 98195, USA}
\altaffiltext{5}{Department of Physics and Astronomy, University of Utah, Salt Lake City, UT 84112, USA}
\altaffiltext{6}{Osservatorio Astronomico di Padova-INAF, Vicolo dell Osservatorio 5, I-35122 Padova, Italy}
\altaffiltext{7}{Space Telescope Science Institute, Baltimore, MD 21218, USA}
\altaffiltext{8}{Hubble Fellow}

\begin{abstract}

We explore the ratio (C/M) of carbon-rich to oxygen-rich thermally pulsing asymptotic giant branch (TP-AGB) stars in the disk of M31 using a combination of moderate-resolution optical spectroscopy from the Spectroscopic Landscape of Andromeda's Stellar Halo (SPLASH) survey and six-filter Hubble Space Telescope photometry from the Panchromatic Hubble Andromeda Treasury (PHAT) survey. Carbon stars were identified spectroscopically. Oxygen-rich M-stars were identified using three different photometric definitions designed to mimic, and thus evaluate, selection techniques common in the literature. We calculate the C/M ratio as a function of galactocentric radius, present-day gas-phase oxygen abundance, stellar metallicity, age (via proxy defined as the ratio of TP-AGB stars to red giant branch, RGB, stars), and mean star formation rate over the last 400~Myr. We find statistically significant correlations between log(C/M) and all parameters. These trends are consistent across different M-star selection methods, though the fiducial values change. Of particular note is our observed relationship between log(C/M) and stellar metallicity, which is fully consistent with the trend seen across Local Group satellite galaxies. The fact that this trend persists in stellar populations with very different star formation histories indicates that the C/M ratio is governed by stellar properties alone.
\end{abstract}


\maketitle

\section{Introduction}
Asymptotic giant branch (AGB) stars are important for understanding galaxies' integrated light and resolved stellar populations. They are major contributors of near-infrared (NIR) flux, contributing $\sim20\%$ of a galaxy's NIR light in the local universe, and up to $70\%$ at high-redshift \citep{Boyer2011a, Melbourne2012, Conroy2013, Melbourne2013, Villaume2015}. They also remain one of the least understood phases of stellar evolution, with outstanding questions regarding calibration of the thermally-pulsating AGB (TP-AGB) phase, dredge-up, opacities and mass loss \citep[e.g.][and references therein]{Marigo2013}. 

AGB stars are broadly characterized by whether their atmospheres contain excess carbon or excess oxygen. Stars with free oxygen (C/O $< 1$) are deemed M-type AGB stars, or M-stars. These stars undergo the third dredge-up (3DU), which brings newly formed carbon and s-process elements to the surface. This pollution alters the atmospheric chemistry, and causes the star to transition first to S-type (C/O$\sim 1$) and finally to C-type (C/O $> 1$). The C-type stars with free carbon are also known as carbon stars. 

Because the transition from M-star to C-star depends on metallicity and stellar mass, the ratio (C/M) of carbon-rich to oxygen-rich AGB stars is a useful tool for studying the evolution of TP-AGB stars, and the galactic environment in which the stars formed. For example, C/M ratios obtained for galaxies throughout the Local Group \citep[e.g.][]{Cioni2003, BattinelliDemers2004, Cioni2008, BattinelliDemers2009, Boyer2013}, have been used to constrain models of AGB stars \citep{Karakas2014} and metallicity gradients of the host galaxies \citep[e.g.][]{Cioni2008, Feast2010}. 

With its complicated star formation history (SFH) and metal-rich environment, M31 is a particularly powerful laboratory for continuing this work. Due to its size and distance, past work on the AGB population in M31 has been limited to relatively shallow ground-based surveys \citep{Brewer1995, Brewer1996, Battinelli2005}, Hubble Space Telescope (HST) pencil beams \citep{Stephens2003, Boyer2013}, or adaptive-optics assisted observation \citep{Davidge2001, Davidge2005}. Fortunately, recent large-scale surveys have opened up the AGB stars of M31 to the same scrutiny as elsewhere in the Local Group.

In this work, we use data from both the Panchromatic Hubble Andromeda Treasury \citep[PHAT;][]{Dalcanton2012} and the Spectroscopic and Photometric Landscape of Andromeda's Stellar Halo \citep[SPLASH;][]{Guhathakurta2006a} surveys. The pairing of HST photometry and Keck spectroscopy across the disk of M31 allows us to not only distinguish carbon-rich and oxygen-rich AGB stars, but also to evaluate the C/M ratio as a function of a wide variety of parameters. We can, for the first time, self-consistently evaluate the C/M ratio across a variety of age, metallicity and SFH environments. 

Section~\ref{data} describes our data, including the surveys from which we obtain our spectra and photometry, and our method of distinguishing between carbon-rich and oxygen-rich AGB stars. Section~\ref{Mdef} examines the effect of different possible AGB classifications on the C/M ratio. Section~\ref{results} presents the resulting C/M ratios as a function of radius, metallicity, and star formation history (SFH). Section~\ref{discussion} discusses these results and possible biases.

\section{Data}\label{data}

\subsection{Photometry}

The photometry used in this work is from the PHAT survey, which resolved $\sim 117$ million stars in the disk of M31 \citep{Williams2014}. Images were taken with UV ($F275W$ and $F336W$), optical ($F475W$ and $F814W$) and NIR ($F110W$ and $F160W$) filters, using HST WFC3/UVIS, ACS/WFC and WFC3/IR, respectively. The photometry was performed using DOLPHOT \citep{Dolphin2002}, and ``good star" (gst) cuts were made using the signal-to-noise (S/N), sharpness, and crowding parameters output by the photometry pipeline. For further details, we refer the reader to \citet{Dalcanton2012} and \citet{Williams2014}.

When comparing the C/M ratio to the stellar metallicity in $\S$~\ref{sec:Z} and the age proxy in $\S$~\ref{sec:Age}, we use the single camera ACS photometry \citep{Dalcanton2012} further corrected for bias, completeness, and foreground extinction by \citet{Gregersen2015} (hereafter G15). Photometric bias is caused by the effect of unresolved and bright stars in crowded regions, which leads stars to appear brighter and with a color closer to the mean color of the region. Completeness, the fraction of stars observed vs. the number of stars present, has a color dependent effect on the stellar population. While completeness does not impact the magnitude of a star, it does impact quantities derived from the population (e.g. median metallicity). To correct for foreground extinction, G15 assume $A_V = 0.17$~mag. For further details on these corrections, we refer the reader to the original paper.

\subsection{Spectroscopy}\label{spec}
The complete spectroscopic observations are documented elsewhere \citep[e.g.][]{Guhathakurta2006a, Gilbert2009, Dorman2012, Gilbert2012, Dorman2015}, but those specific to the disk are briefly summarized here. 

Our spectroscopic targets were selected from optical photometry. The majority ($\sim 63\%$) were targeted using HST photometry from PHAT. The rest were targeted using CFHT $i'$ photometry \citep[discussed in][]{Gilbert2012}. In all cases, stars were selected to be isolated, without close, bright neighbors \citep[see][for details]{Dorman2012}.

The disk dataset contains 10,619 optical spectra taken with the DEIMOS spectrograph \citep{Faber2003} on the Keck II 10-m telescope. Approximately half (5323) of these spectra were observed with the 1200 line~mm$^{-1}$ grating, which has a dispersion of 0.33\AA~pixel$^{-1}$ and a central wavelength of $7760$\AA. The rest (5296) were taken with the 600 line~mm$^{-1}$ grating, which has a dispersion of 0.65\AA~pixel$^{-1}$ and a central wavelength of 7000\AA. 

The spectra were reduced using the \texttt{spec2d} and \texttt{spec1d} software modified for the SPLASH survey \citep{Cooper2012, Newman2013}. The \texttt{spec2d} routine firsts extracts a one-dimensional spectrum from the two-dimensional spectral data, and the \texttt{spec1d} routine determines the redshift of this spectrum by cross-correlating with a series of templates. This cross-correlation also serves to distinguish galaxies and failed observations that returned no flux. 

We refer to these 10619 spectra as our ``science spectra," and they have a median S/N of 3.4 pix$^{-1}$.

\subsection{Spectroscopic Templates}

\begin{figure}
\begin{center}
\includegraphics[width = 3.5in]{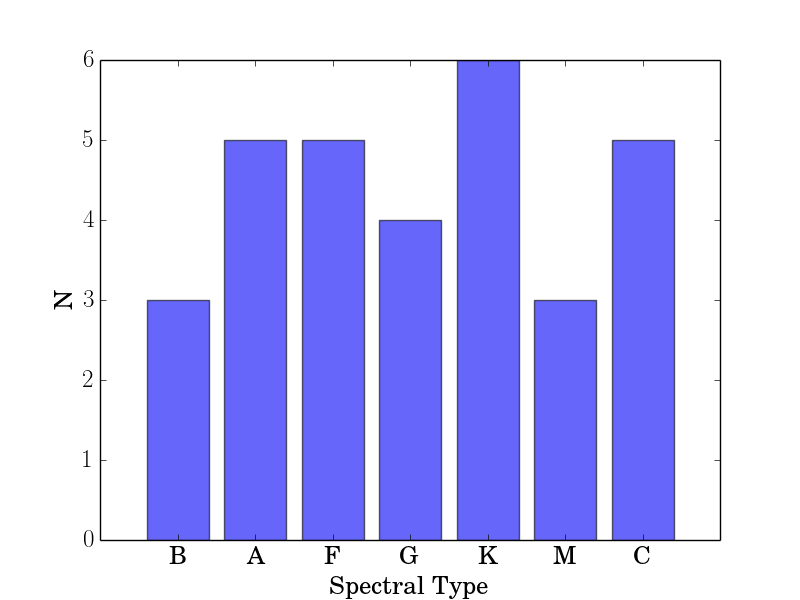}
\caption{Distribution of spectral types represented in our suite of high S/N Milky Way template spectra. ``C" denotes carbon stars.}
\label{fig:temptypes}
\end{center}
\end{figure}

To complement the Milky Way spectroscopic templates included in the \texttt{spec2d} pipeline, we observed a series of Milky Way carbon stars to use as radial velocity and spectral type templates. The stars were selected from the SIMBAD database \citep{Wenger2000} based on magnitude, position, and spectral type. Their apparent magnitudes are in the range $7.74 > V > 6.21$ and their spectral types span C5 to C8 based on the Keenan and Morgan classification scheme \citep{KeenanMorgan1941}.

The observations were carried out using the 600 line~mm$^{-1}$ DEIMOS grating centered at 7000\AA~and the GG455 filter to block shorter wavelength light. We use the long-slit mask LVMslits with a 0.8" wide slit. This instrumental configuration provides a wavelength coverage of 4800-9500\AA~with a spectral resolution of 0.65\AA~pixel$^{-1}$. The raw two-dimensional spectra were reduced and extracted in the same manner as the science spectra.

Our final suite of spectroscopic templates contains 31 stars with a median S/N of 43 pix$^{-1}$, significantly higher than our 10619 M31 spectra. The distribution of template spectral types is shown in Figure~\ref{fig:temptypes}. When referring to these spectra, we will use the term``template spectra" to distinguish them from the ``science spectra."

\subsection{AGB Identification}\label{phatID}

\begin{figure*}
\begin{center}
\includegraphics[width = 7in]{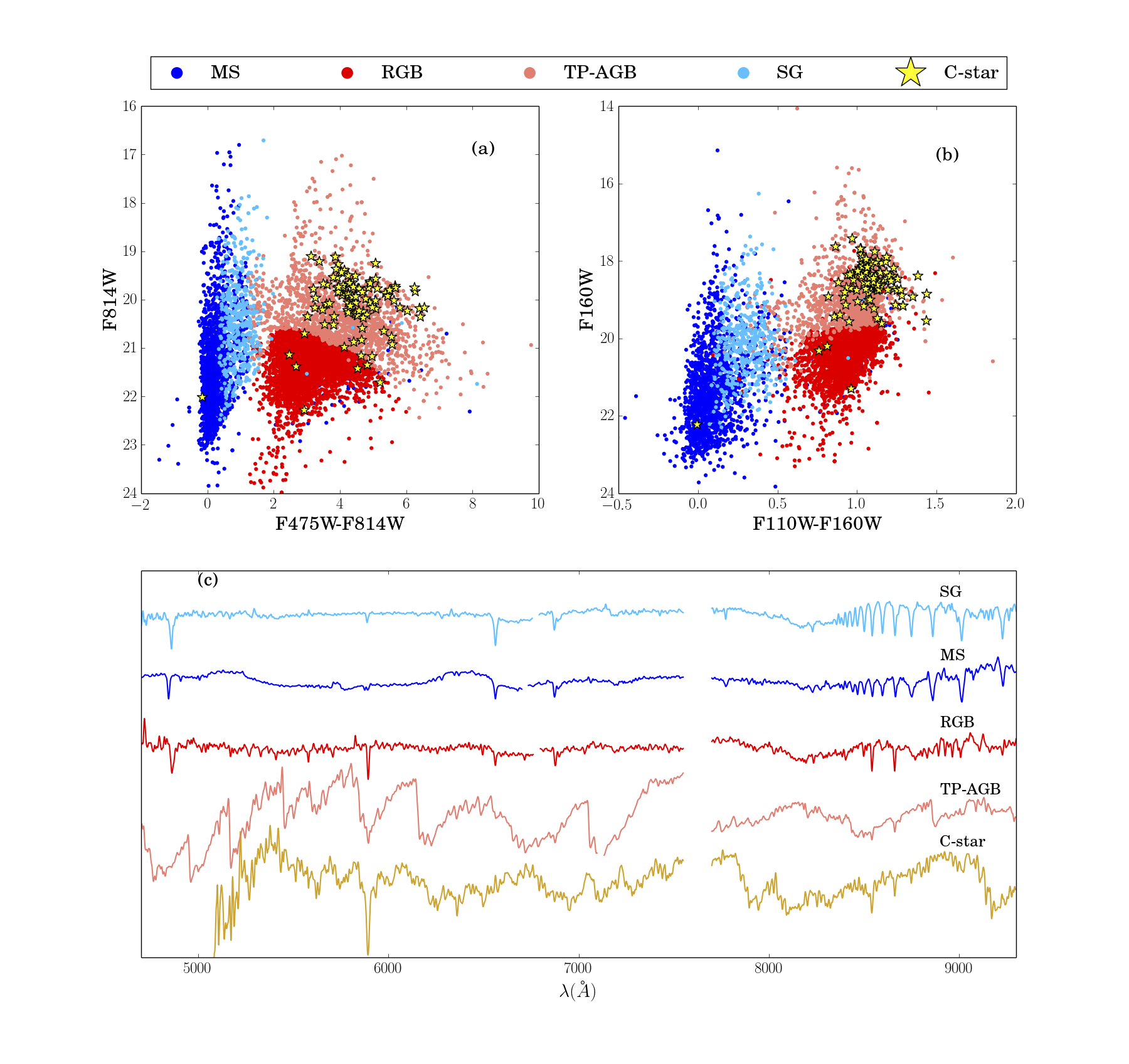}
\caption{Panel (a) shows an optical ($F814W$ vs. $F475W-F814W$) CMD of stars in the SPLASH sample, color-coded by evolutionary stage as determined from the photometry. Main sequence stars are plotted in blue, super giants in cyan, RGB stars in red, and AGB stars in pink. We show those stars identified as spectroscopically as carbon stars as yellow stars. Panel (b) shows the same stars on a NIR ($F160W$ vs. $F110W-F160$) CMD. Panel (c) shows a representative DEIMOS spectrum of each of the color-coded populations in the top panels. Spectra are normalized, smoothed by a Gaussian with $\sigma\sim3$ pixels, and plotted with a vertical offset. We also mask the the telluric A band at 7600\AA. }
\label{fig:starID}
\end{center}
\end{figure*}

We use position on the PHAT color-magnitude diagrams (CMDs) identify stars at different evolutionary stages. The regions discussed in this section are illustrated in Figure~\ref{fig:starID}.

Stars on the red giant branch (RGB) were identified using optical ($F475W$ and $F814W$) Padova PARSEC1.2s isochrones \citep{Bressan2012}. The RGB boundaries, and most importantly the tip of the red giant branch (TRGB), were defined using isochrones with $-2.1 < {\rm [M/H]} < 0.6$ and a fiducial age of 10~Gyr. We assumed a foreground reddening of $A_V = 0.17$~mag and a distance to M31 of $776 \pm 18$~kpc \citep{Dalcanton2012}. For an example of the RGB footprint, see Figure~1 of G15. Super giants (SG) were identified as a distinct branch in $F336W-F475W$ vs. $F110W - F160W$ color-color space. All objects not classified as SG or RGB were classified based on optical color and magnitude. Those stars with $F475W - F814W < 1.2$ are assumed to be main sequence (MS) stars, and those stars brighter than the TRGB with $F475W - F814W > 1.2$ are assumed to be AGB stars. These designations are purely photometric, with no reliance on spectral features. Our spectroscopic sample is dominated by RGB stars (5680), with 2489 MS stars and 1867 AGB stars. The remaining 583 objects are classified as either SG or PNe.

These designations are such that the bluest AGB stars may be bright supergiants and/or core helium-burning stars. We cannot distinguish between these objects and early-type AGB stars photometrically or spectroscopically at our resolution \citep{Melbourne2012, Dalcanton2012}. Alternatively, some may be the reddest (and relatively few) Milky Way stars that are observed at very red colors \citep{Williams2014}. In addition, the distinction between RGB and AGB stars means our AGB sample is dominated by TP-AGB stars. 

\subsection{Carbon Star Identification}\label{cid}

\begin{figure}[t!]
\begin{center}
\includegraphics[width=3.5in]{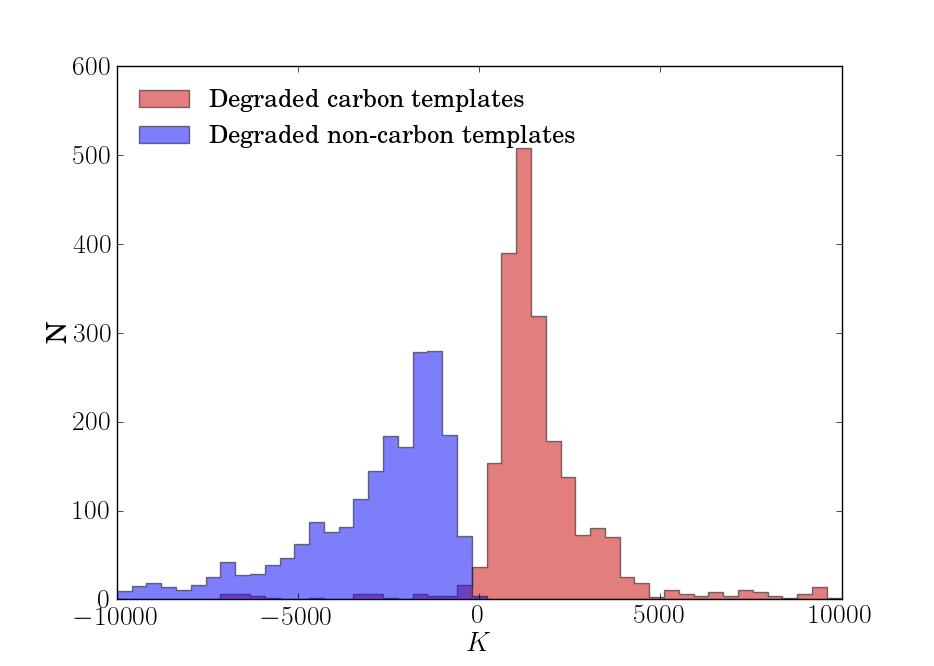}
\caption{Evaluation of the $K$-statistic based carbon star identification method. The solid histograms show the distribution of $K$ values for our degraded templates with S/N$>2$. For ease of visualization, we truncate the x-axis, and note that templates degraded to a relatively high S/N often have $K$ values beyond these bounds.}
\label{fig:Ktest}
\end{center}
\end{figure}

Our primary method of identifying carbon stars compares each science spectrum to our suite of 31 Milky Way template stars. To compute whether a science spectrum is best fit by a carbon template or a non-carbon template, we use a simple classification statistic $K$ defined as:
 \begin{equation}
 K = {\rm min}\left(\chi^2(\theta_{NC})\right)-{\rm min}\left(\chi^2(\theta_C)\right)
 \end{equation}
 
\noindent where $\theta_{NC}$ denotes the suite of non-carbon template stars, and $\theta_{C}$ is the suite of carbon template stars. $K$ is the difference between the $\chi^2$ statistic for the best fitting non-carbon template and the $\chi^2$ statistic for the best fitting carbon template. By this definition, those stars with $K > 0$ are likely carbon stars. This metric is loosely based on the ratio of evidence, or Bayes Factor, with a minimization rather than a sum.

We verify this method by degrading the S/N of each template to mimic the range of S/N present in our data, and then computing the degraded template's $K$ value by comparing it to the remaining 30 templates. This process is repeated until there are $\sim3000$ degraded carbon templates and $\sim3000$ degraded non-carbon templates. Figure~\ref{fig:Ktest} shows the distribution of $K$ for our degradation tests. They indicate that this method is $\sim 88\%$ accurate across the full range of S/N, and $\sim 95\%$ accurate for those stars with S/N $\ge 2$~pix$^{-1}$.

\begin{figure}[h!]
\begin{center}
\includegraphics[width = 4in]{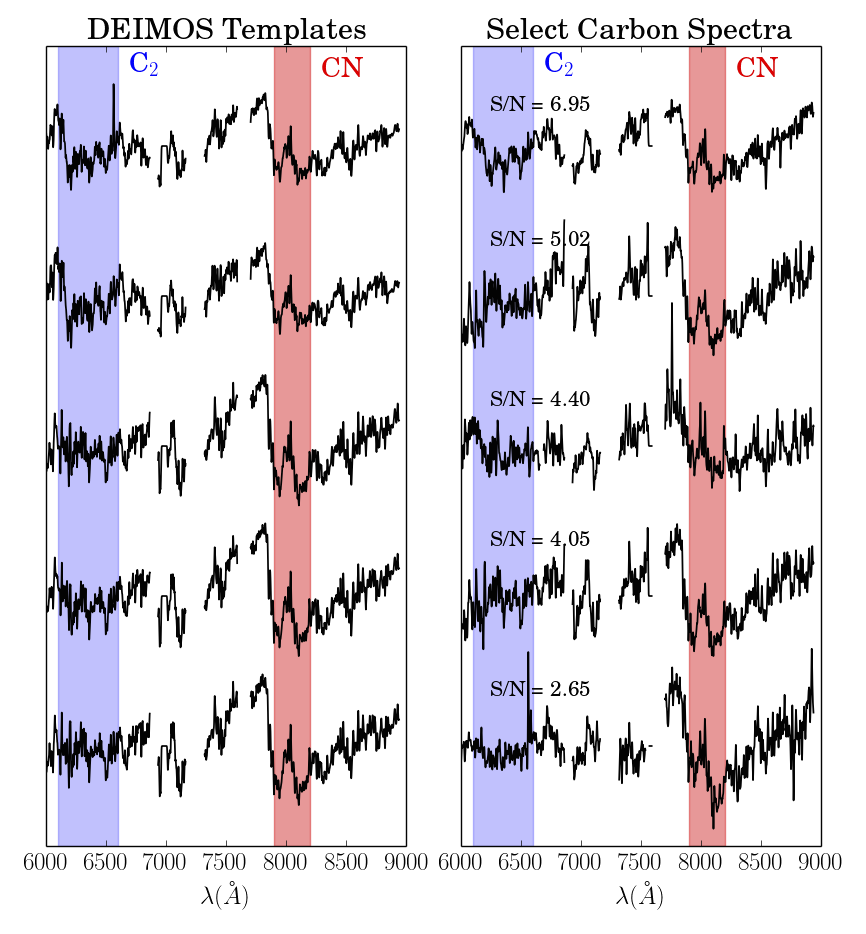}
\caption{The left panel shows the five MW carbon stars used as templates for identification via $K$-statistic. The right panel shows five M31 carbon stars chosen to demonstrate a range of S/N. Spectra are smoothed by a Gaussian with $\sigma\sim3$ pixels, and the the telluric A band at 7600\AA~is masked. The science spectra are labeled with their S/N pix$^{-1}$. On both panels we have labeled the region where C$_2$ is prominent in blue ($\sim 6100-6600$ \AA), and the region where CN is prominent in red $\sim 7900-8200$\AA). These are the features we look for during the visual inspection that follows any automated carbon star detection method. }
\label{fig:temp spec}
\end{center}
\end{figure}

We apply our classification statistic to the 7903 stellar spectra with S/N pix$^{-1}>2$, and visually inspect the spectra that return $K > 0$. Specifically, we look for the CN features at $\sim7900$\AA~and C$_2$ features between 6100 and 6600\AA, shown in Figure~\ref{fig:temp spec} for the five carbon templates and five randomly chosen carbon science spectra. We find a total of 94 carbon stars. There are 316 objects whose spectra return $K > 0$ but upon inspection do not have the carbon features highlighted in Figure~\ref{fig:temp spec}. This is a higher false-positive rate than our tests predicted, because our tests did not simulate the bad sky subtraction or untrustworthy radial velocity measurements common for a multi-slit spectrograph.

\begin{figure}[h!]
\begin{center}
\includegraphics[width=3.5in]{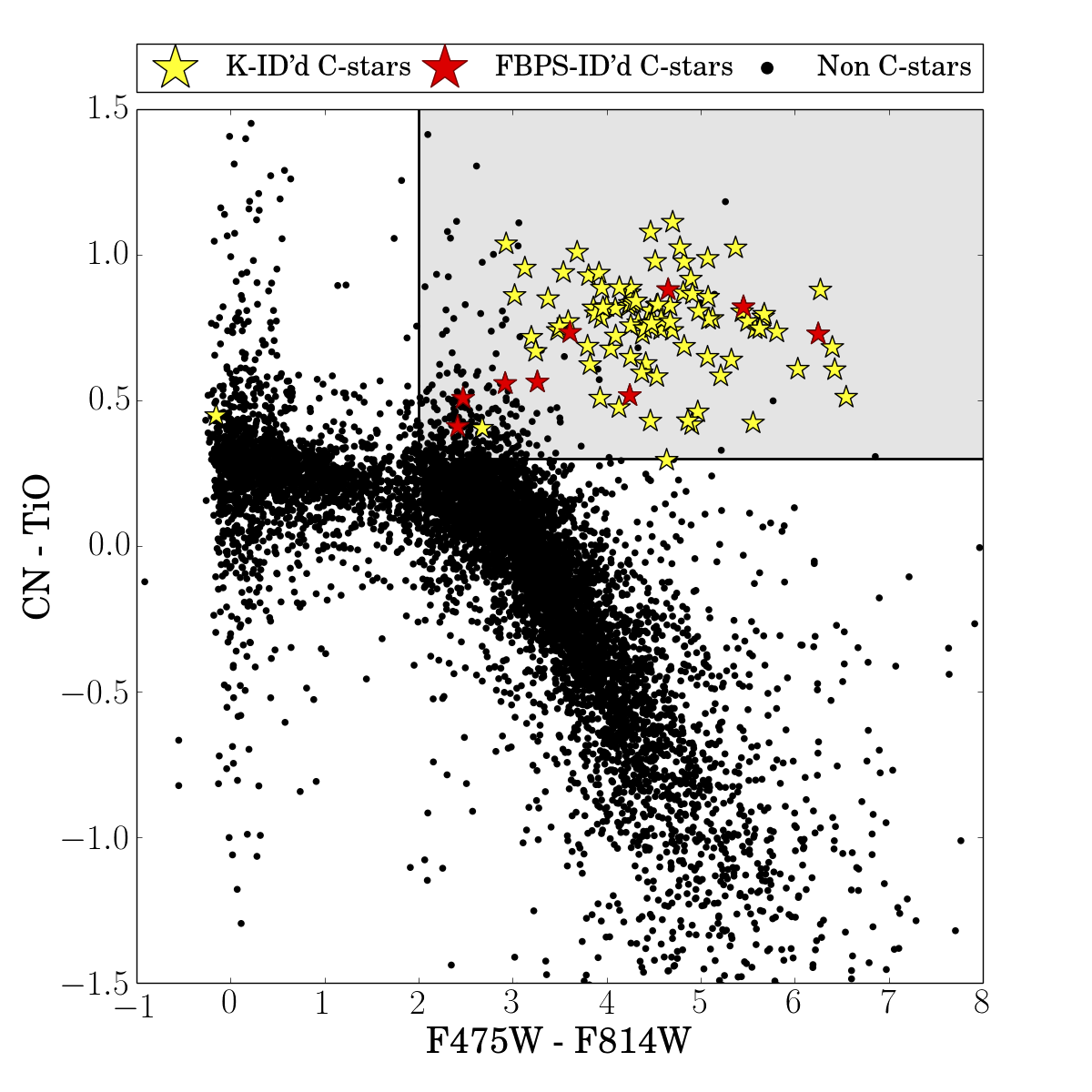}
\caption{FBPS Color-color diagram ($F475W-F814W$ vs. synthetic CN$-$TiO) used to identify potential carbon stars. The grey shaded region is defined by the bounds $F475W - F814W > 2.0$ and CN--TiO$>0.3$, and represents where carbon stars are likely to lie. Carbon stars identified by the $K$ classification statistic are plotted as yellow stars, and carbon stars identified using this FBPS photometry (plus visual inspection of the spectra) are plotted as red stars. All other stars are plotted as black points.}
\label{fig:FBPS}
\end{center}
\end{figure}

To catch carbon stars that may have escaped identification by the $K$-statistic, we run a separate identification using synthetic narrow-band photometry modeled on the four band photometry system (FBPS). FBPS has been used extensively throughout the Local Group \citep[e.g.][]{Nowotny2003, Battinelli2004a, Battinelli2004, Wing2007, Battinelli2009}. FBPS uses broad band color (i.e. $R-I$, $V-I$) in conjunction with the color defined by the narrow band CN and TiO filters (centered at 8120.5\AA~and 7778.4\AA, respectively) to separate C- and O-rich AGB stars.  It is one of two preferred photometric techniques for this separation \citep[the other being NIR $J$ and $K_s$ photometry, whose reliability has recently been called into question;][]{Menzies2015}.

To compute synthetic CN and TiO magnitudes from our spectra, we perform a first-order flux calibration. We create a transmission curve by dividing our MW template spectrum of HD52005 with the fully flux-calibrated spectrum of HD52005 from the X-shooter Spectral Library \citep{Chen2014}, and use this transmission curve to correct the shape of our spectra. We then weight the spectra by the CFHT/CFH12k CN and TiO throughput curves to generate synthetic photometry.\footnote{http://svo2.cab.inta-csic.es/svo/theory/fps3/} We use $F475W - F814W$ as our broad band color. 

Because this method is computationally inexpensive, we run it on all stellar spectra to catch any carbon stars whose S/N excluded them from the $K$-statistic method. The resulting FBPS diagram is shown in Figure~\ref{fig:FBPS}. We define the region of likely carbon stars using the position of carbon stars identified by the $K$ classification statistic. The bounds are $F475W - F814W> 2.0$ and CN--TiO $> 0.3$, which is fully consistent with the bounds used throughout the literature. We visually inspect the spectra of each star in the bounded region, and find an additional 9 carbon stars. 

In total, we find 103 carbon stars in our sample of stars in the disk of M31. By using this combination of two automated-detection techniques plus visual inspection, we are confident that these 103 stars represent the vast majority of carbon stars in our spectroscopic sample.  

While the majority of these carbon stars lie predictably above the TRGB, four are found in other areas of the CMD (see Figure~\ref{fig:starID}). Three have colors and magnitudes matching the RGB population, and one lies in the MS. To verify that these stars are members of the M31 disk rather than foreground MW carbon dwarfs, we compute their line-of-sight (LOS) velocities. We convert each star's redshift measurement to a velocity, then apply a telluric A-band correction and a correction for heliocentric motion on the date the spectrum was obtained \citep[for further details see][and references therein]{Dorman2012}. These four carbon stars have LOS velocities between $-85$~km/s and $-300$~km/s, making them likely M31 members. Their unusual colors and magnitudes are likely due to a mismatch between the SPLASH and PHAT catalogs.

\section{Defining M}\label{Mdef}
Our spectroscopic identification gives us an uncontaminated sample of unambiguous carbon stars. There exists no equivalent spectroscopic selection that can separate M-type TP-AGB stars from the RGB. And while the C/M ratio has been computed across the Local Group, there is no standard photometric definition for M-stars. Even within M31, multiple criteria have been devised to isolate M-stars based on available data. We examine three criteria based on the work by \citet{Boyer2013} (hereafter B13), \citet{Brewer1995} (hereafter B95), and \citet{BattinelliDemers2005} (hereafter BD05). Our goal is not to define a single, ``correct" method, but rather to investigate the differences in results returned by each of these methods.

\subsection{Boyer et~al. 2013}
The first definition of M that we apply to our data is one modified from B13. As they use WFC3/IR medium-band filters to discriminate between C- and M-type TP-AGB stars, our dataset is certainly not identical. IR filters are far more sensitive to dust-enshrouded stars than optical filters, and so the B13 sample will be substantially more complete. However, they make two major assumptions that we can emulate: TP-AGB stars are brighter than TRGB, and all AGB stars not classified as C-stars are M-stars. Their definition of the TRGB also comes from PHAT photometry, so this identification of AGB stars matches our own. If we too take all non-carbon AGB stars to be M-type, we get a sample of 1605 stars from our original sample of 10619 spectra.

Whenever this particular sample is used, it will be denoted by a red diamond and will be referred to as the B13 method. 

\begin{figure}[h!]
\begin{center}
\includegraphics[width = 2.8in]{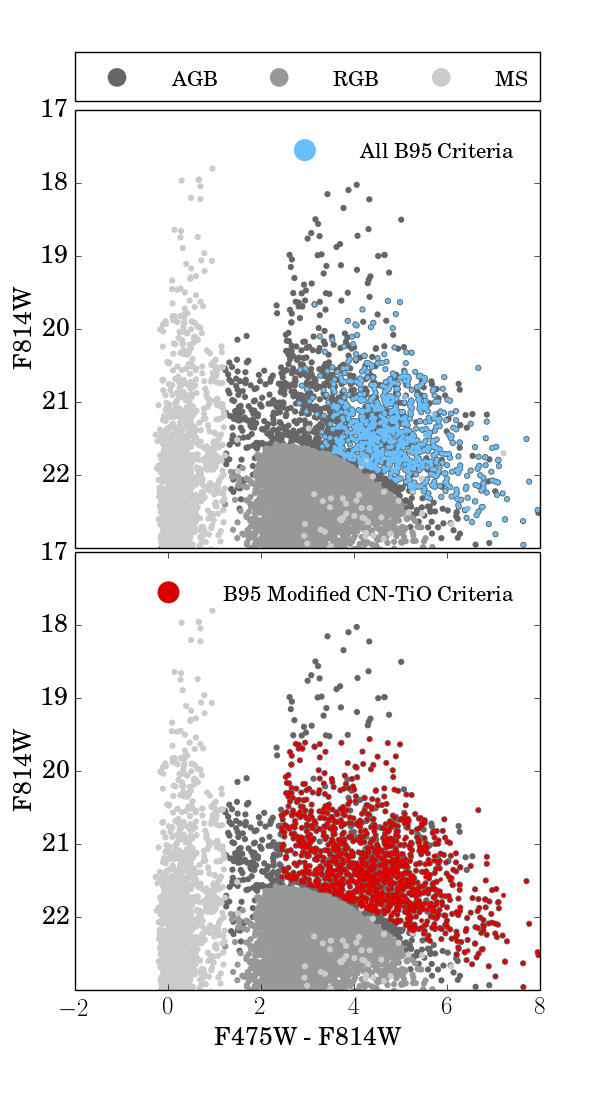}
\caption{The results of applying the FBPS-based M-giant selection criteria of Brewer et.~al 1995 to our spectroscopic sample. The top panel includes all selection criteria; $V-I > 1.8$, CN$-$TiO$~< -0.2$, $I > 18.5$, and $M_{\rm bol} < -3.5$. $M_{\rm bol} = I + 0.3 + 0.34(V-I)-0.14(V-I)^2 - (m-M)_0$. The bottom panel includes all broad-band color and magnitude cuts but replaces the CN$-$TiO with the criterium that a star not be a spectroscopically-identified carbon star. AGB, RGB, and MS stars (denoted in grays) come from PHAT CMD-based identification. We adopt the cuts in the top panel for the B95 selection used in the rest of this paper.}
\label{fig:B95}
\end{center}
\end{figure}

\subsection{Brewer et~al. 1995}
B95 is one of the few large-scale AGB surveys of M31. The authors used FBPS in five regions located along the southern major axis. They define M by the following criteria: $V-I > 1.8$, CN$-$TiO$~< -0.2$, $I > 18.5$, and M$_{bol} < -3.5$ (where M$_{bol} = I + 0.3 + 0.34(V-I)-0.14(V-I)^2 - (m-M)_0$). They assume a distance modulus of 24.41, adopted from \citet{Freedman1990}. 

To replicate these criteria for our sample, we convert our $F475W$ and $F814W$ magnitudes to $V$ and $I$. To convert from $F814W$ to $I$ we use the 97 M-giant spectral templates presented by \citet{Fluks1994}. We convolve the Fluks et~al. spectra with the HST/ACS $F814W$ bandpass and the CFHT/CFH12k $I$ bandpass. The offset between $F814W$ and $I$ is a function of color/spectral type. We fit a fourth-order polynomial to this relationship to determine a conversion function, getting residuals less than 0.002 magnitudes. To derive $V-I$ we use the published transformations from \citet{Sirianni2005}.

The application of B95 selection criteria is shown in Figure~\ref{fig:B95}. When we apply all selection criteria from B95, our final sample contains 843 M-stars (shown in blue in the top panel of Fig~\ref{fig:B95}). The sample is relatively uncontaminated, with only 2 stars having been identified by PHAT as RGB stars rather than AGB stars. However, it misses the bluest AGB stars ($F475W-F814W < 3$), the brightest AGB stars, and those stars close to the TRGB. The latter effect grows more pronounced at redder colors, and thus the more metal-rich populations. If we were to remove the CN$-$TiO criterium and replace it with the requirement that no star be a spectroscopically identified carbon star (see \S~\ref{cid}), the sample would increase to 1218 (shown in red in the bottom panel of Fig~\ref{fig:B95}), decreasing the resulting C/M ratio. This expands the sample to include many of the bluer AGB stars left out of the B95 sample. 

In the remainder of the paper, we will use the unmodified B95 selection of 843 stars. Whenever this sample is used it will be denoted by a blue square and will be referred to as the B95 method.

\subsection{Battinelli \& Demers 2005}
A common way to express the C/M ratio in the Local Group is C/M0+; i.e. the M-type AGB stars counted are those with a spectral type of M0 or later. This spectral-type selection is often defined photometrically using a color cut (e.g. $R-I > 0.9$). We can apply a color cut and determine spectral types from the spectra themselves. However, we must first note that the PHAT photometric observations and SPLASH spectroscopic observations were taken several months (or in some cases years) apart. Since a significant fraction of TP-AGB stars are long-period variables of sizable amplitude, the measured PHAT colors may not reflect the colors of the star at the time the spectrum was taken.

\begin{figure}[h!]
\begin{center}
\includegraphics[width = 3in]{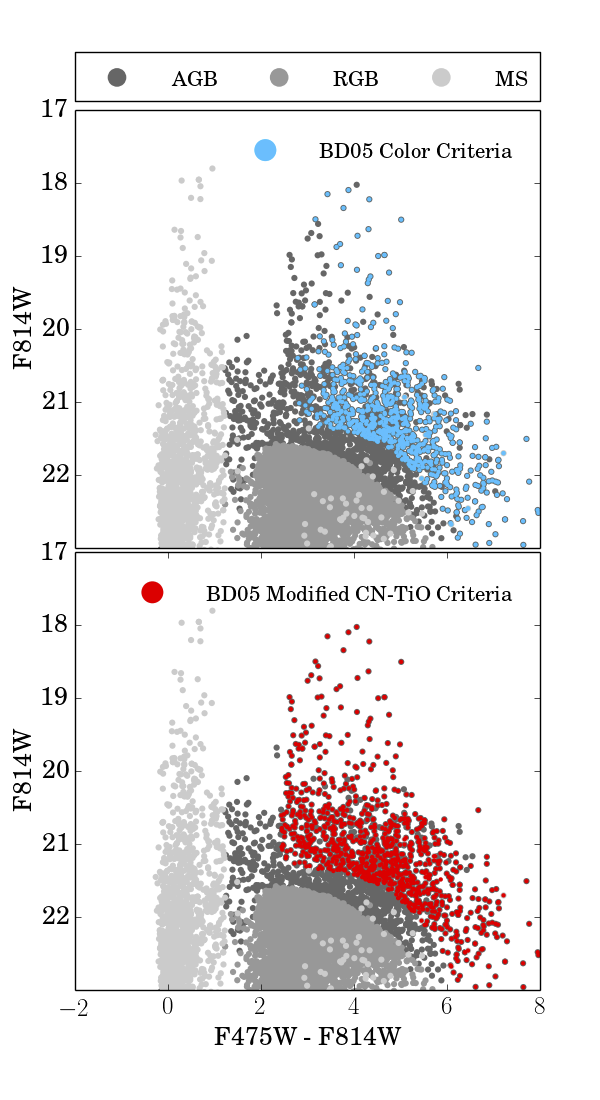}
\caption{The results of applying the FBPS-based M-giant selection criteria of Battinelli \& Demers~2005 to our spectroscopic sample. The top panel includes all BD05 selection criteria; $R-I > 0.9$, CN$-$TiO$~< 0$, and $M_{\rm bol}<-3.5$. Here, $M_{\rm bol} = I + 1.7103 - 2.2968 (R-I) + 1.66464 (R-I)^2 - 0.43399 (R-I)^3 - (m-M)_0$. The bottom panel replaces the CN$-$TiO criteria with the requirement that a star not be a spectroscopically identified carbon star. AGB, RGB, and MS stars (denoted in grays) come from PHAT CMD-based identification. We adopt the cuts in the top panel for the B05 selection used in the rest of this paper.}
\label{fig:B051}
\end{center}
\end{figure}

BD05 define M0+ to be those stars with $R-I > 0.9$ and CN--TiO$~< 0$. This broad-band color criterium matches that used by B95, as $R-I = 0.9$ corresponds to $V-I = 1.8$. To distinguish M-type AGB stars, they use a bolometric magnitude limit of $M_{\rm bol} < -3.5$, where they adopt a slightly different definition of $M_{\rm bol}$ than B95: $M_{\rm bol} = I + 1.7103 - 2.2968 (R-I) + 1.66464 (R-I)^2 - 0.43399 (R-I)^3 - (m-M)_0$. BD05 adopt a distance modulus $(m-M)_0 = 24.41$. To apply these same criteria we first transform $F475W$ and $F814W$ magnitudes into $I$ (using the method outlined in the previous section), and then derive $R$ using the transformations by \citep{Sirianni2005}. The resulting sample is shown in blue in the top panel of Figure~\ref{fig:B051}. It contains only 736 stars, as the sample stops half a magnitude above the TRGB. This sample also omits the bluest AGB stars, but unlike B95, it does contain the brightest. As with the B95 criteria, if we were to replace the CN$-$TiO limit with the requirement that no star be a spectroscopically identified carbon star the sample would increase to 960 stars (shown in red in the bottom panel of Fig~\ref{fig:B051}), decreasing the resulting C/M ratio. As before, this would extend the sample in the blue but would have no effect on the proximity of the sample to the TRGB.

\begin{figure}[b!]
\begin{center}
\includegraphics[width = 3.5in]{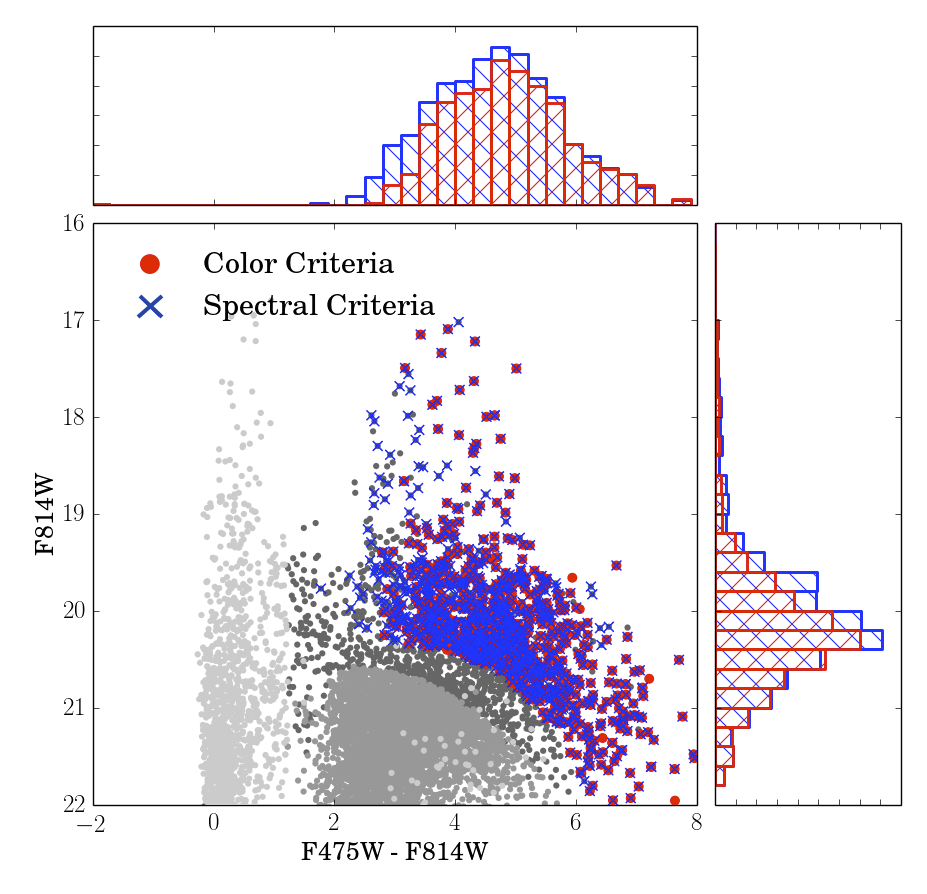}
\caption{The result of a spectrum-based M-giant selection criterion applied to our sample. As in Figures~\ref{fig:B051} and \ref{fig:B95}, AGB, RGB, and MS stars are in greys. Yellow crosses represent those stars with spectral types of M0 or later, as determined by the indices from \citet{Fluks1994}}
\label{fig:B052}
\end{center}
\end{figure}

Since we have spectra in addition to photometry, we can also spectroscopically determine which stars have a subtype of M0 or later. We use the TiO-based spectral parameters, $S_{1/2, S_p}, S_{1/3, S_p},$ and $S_{2/3, S_p}$ defined in \citet{Fluks1994}. These three parameters are calculated by integrating over the various TiO absorption features, and are monotonic functions of spectral type. Stars with subtype M0 or later have $S_{1/2, S_p}  < 0.517, S_{1/3, S_p} < 0.754$, and $S_{2/3, S_p} < 1.458$. There is no spectroscopic distinction between AGB and RGB stars at this resolution, so we again apply the bolometric magnitude limit from BD05. This returns a sample of 773 M-stars, shown in yellow on Figure~\ref{fig:B052}. The 736 M-stars identified by the photometric M0+ criteria are plotted in red.

The sample of M-stars identified as M0+ by spectroscopic criteria are almost identical to those identified by the photometric criteria. The top and right panels of Figure~\ref{fig:B052} shows the marginal distributions of both the spectroscopic and photometric sample in color and magnitude. The color criteria are slightly less sensitive to fainter and bluer stars ($F814W  > 18$). 

In the remainder of the paper, we will use the unmodified BD05 photometric selection of 736 stars. Whenever this sample is used it will be denoted by a green circle and will be referred to as the BD05 method.

\section{C/M ratio across M31}\label{results}

The broad range of environments in M31 mean that we can compute the C/M ratio as a function of a variety of properties. In this section we look at the C/M ratio as a function of galactocentric radius ($R_{\rm g}$, \S~\ref{rad}), metallicity (\S~\ref{results:metallicity}), and SFH (\S~\ref{sec:Age}) in spatial bins. 

\begin{figure}
\begin{center}
\includegraphics[width = 4in]{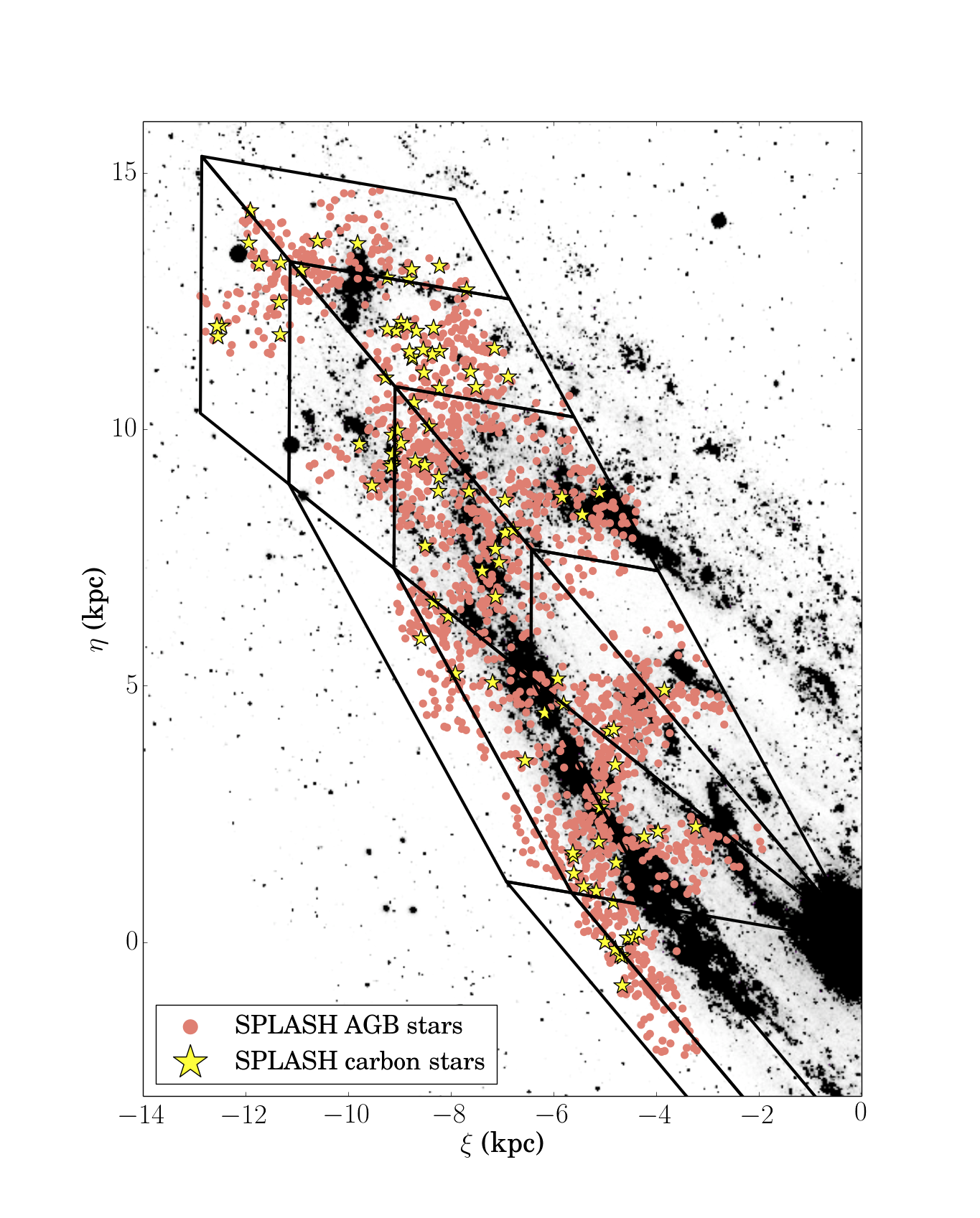}
\caption{Positions of all AGB stars in the SPLASH sample (pink points), with carbon stars (yellow stars), overlaid on a GALEX image of M31. The spatial bins used throughout \S\ref{rad}, \S\ref{results:metallicity} and \S\ref{sec:Age} are marked with black lines. Apparent structure in the positions of the AGB stars is due to the SPLASH selection function, and is not a true physical effect.}
\label{fig:map}
\end{center}
\end{figure}

We create the spatial bins used throughout this section beginning with equal-area bins on a flat circle with a radius of 20~kpc (roughly the outer limit of our data). We then incline the bins to match M31's inclination of $74^\circ$ \citep{Barmby2006} and tilt the bins to match M31's position angle of $50^\circ$ (determined empirically). The final positions of these bins are shown along with the positions of our C- and M-stars in Figure~\ref{fig:map}. 

\subsection{As a function of galactocentric radius}\label{rad}

\begin{figure}[t!]
\begin{center}
\includegraphics[width = 3.5in]{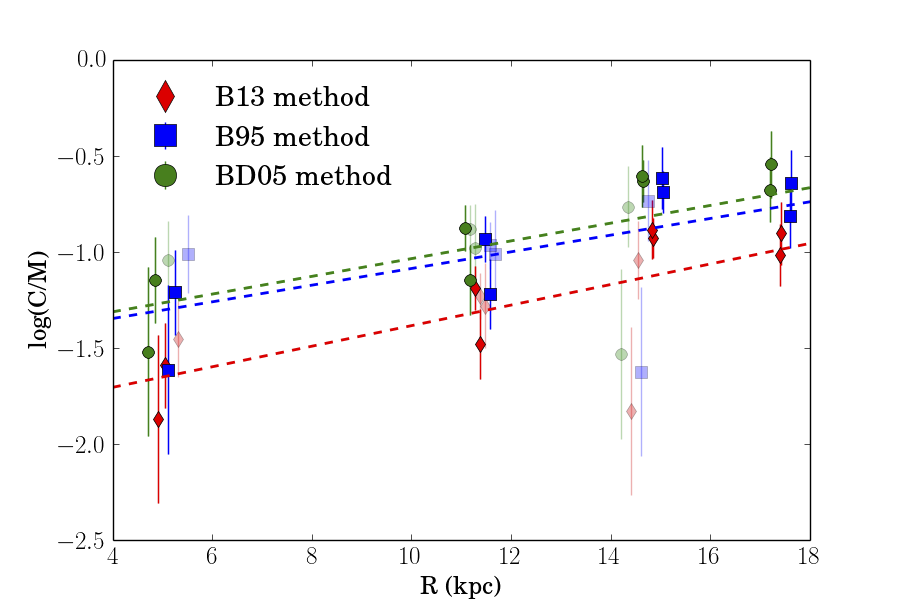}
\caption{Log(C/M) ratio in bins of $R_{\rm g}$. The radii are those of the mean stellar distance in each bin. Red diamonds represent points calculated using the B13-based definition of M, blue squares represent the B95-based definition of M, and green diamonds represent the BD05-based definition of M. For ease of visualization, B95 and BD05 points are artificially offset from the B13 points by 0.2 and -0.2 kpc, respectively. Bins along the major axis are shown as solid points and bins along the minor axis are shown as transparent points.}
\label{fig:withRad}
\end{center}
\end{figure}

Using the bins shown in Figure~\ref{fig:map}, we determine the change in the C/M ratio as a function of radius. We compute the center radius of each bin by taking the average $R_{\rm g}$ of all PHAT stars within that bin. When computing this average, we leave out PHAT ``Brick 1."  Brick 1 covers the extremely crowded bulge region, extending only $\sim2.7$~kpc from the center of M31, where we have no spectroscopic coverage. Leaving those stars out of our average ensures that the resulting distances represent the regions in which our AGB stars lie. The resulting C/M ratios for each of the selection criteria discussed in \S\ref{Mdef} are shown in Figure~\ref{fig:withRad}.

There is a clear positive trend. A weighted least squares fit to the eight bins along major axis bins determines the following gradients:
\begin{flalign*}
{\rm log(C/M_{B13})} &= (0.06 \! \pm \! 0.016)\! \times \! R_{\rm g}\! - \! (1.92 \! \pm \! 0.22) & \\
{\rm log(C/M_{B95})} &= (0.05 \! \pm \! 0.016)\! \times \! R_{\rm g}\! - \! (1.56 \! \pm \! 0.23) \\
{\rm log(C/M_{BD05})} &= (0.06 \! \pm \! 0.017)\! \times \! R_{\rm g}\! - \! (1.53 \! \pm \! 0.23) 
\end{flalign*}

This trend is likely to be driven not by the radius itself, but by other quantities that vary with radius. We consider correlations with these quantities in the following two sections.

\subsection{As a function of metallicity}\label{results:metallicity}
We are in a unique position to compute the C/M ratio as a function of metallicity calculated using different techniques and for different tracer populations. In this section we focus on gas-phase metallicity, as determined from H{\sc II} regions, and on photometrically derived stellar metallicity. 

\subsubsection{Gas-phase metallicity}\label{oh}

\begin{figure}[b!]
\begin{center}
\includegraphics[width = 3.5in]{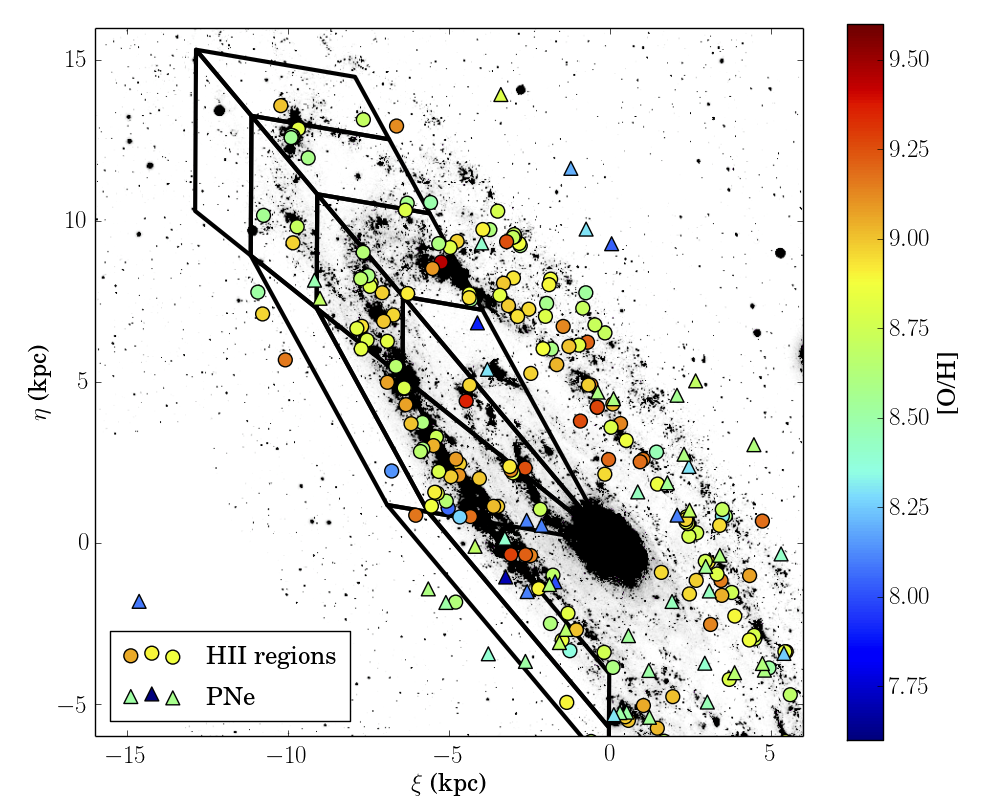}
\caption{Positions of H{\sc II} regions (circles) and PNe (triangles) from Sanders et~al. 2012, with respect to our spatial bins (black lines). Points are colored by their oxygen abundance.}
\label{fig:OHmap}
\end{center}
\end{figure}

We first look at the gas-phase metallicities from \citet{Sanders2012}. The authors measure the oxygen abundance ([O/H]) in H{\sc II} regions using strong-line diagnostics from \citet{Zaritsky1994}, \citet{Kewley2002}, \citet{Nagao2006}, and \citet{Pilyugin2005}. They find the first three methods very consistent, but note that the method from \citet{Pilyugin2005} has a median offset $\sim5\times$ that of the others. In light of this, when compiling the data from \citet{Sanders2012} we disregard the measurements from \citet{Pilyugin2005} and take a weighted average of the rest. The positions of the H{\sc II} regions are shown with respect to our data in Figure~\ref{fig:OHmap}.

We compute the C/M ratio as a function of gas-phase [O/H] using the bins shown in Figure~\ref{fig:map}. The [O/H] value for each bin is defined as the [O/H] value of the H{\sc II} region closest to the average position of all PHAT stars within the bin. We chose this method rather than an interpolation because \citet{Sanders2012} find that the [O/H] distribution is quite clumpy. While there is a global metallicity gradient of $-0.0195 \pm 0.0055$ dex kpc$^{-1}$, H{\sc II} regions even $\sim2$~kpc apart may have dramatically different abundances. 

\begin{figure*}
\begin{center}
\includegraphics[width = 6.5in]{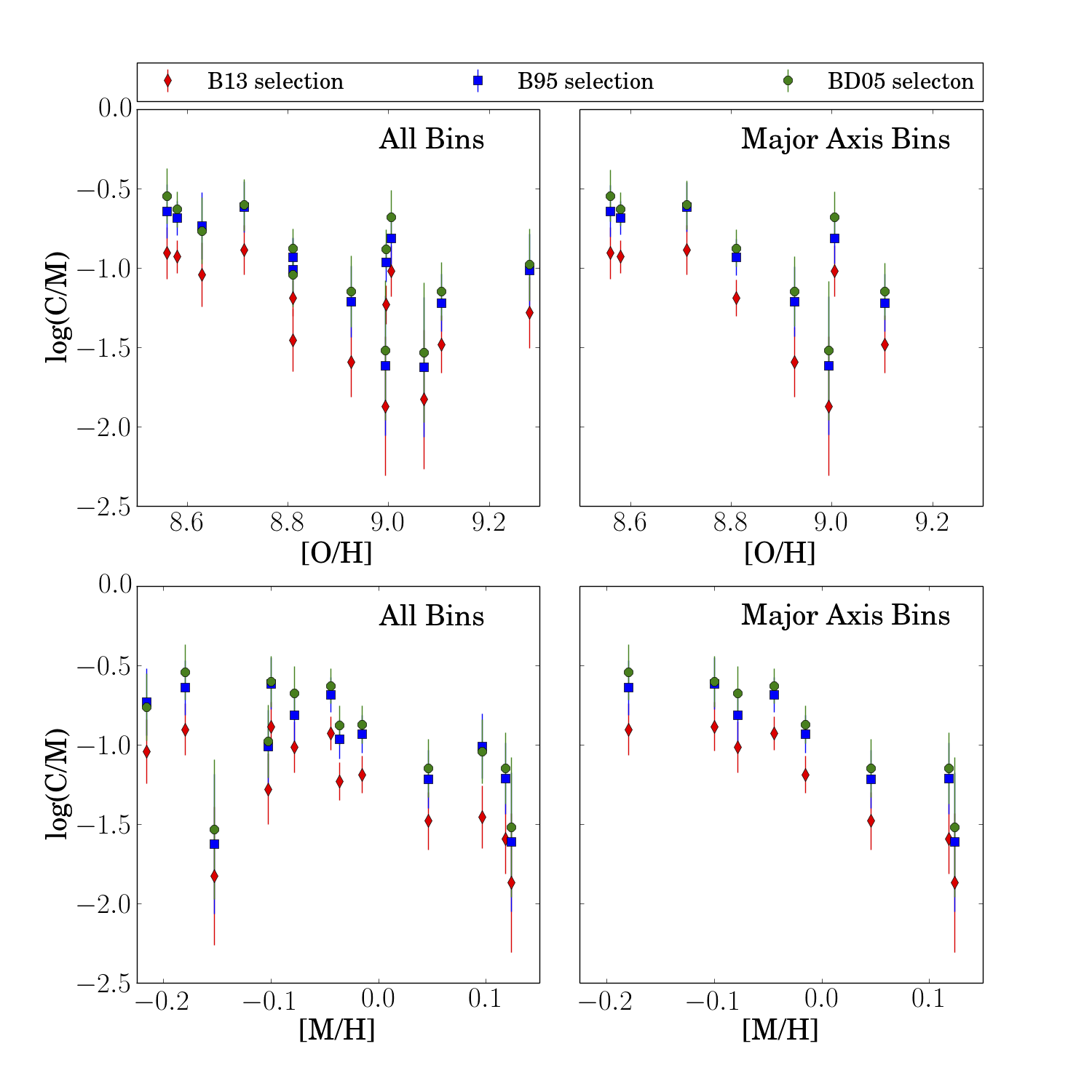}
\caption{Log(C/M) vs various properties of the M31 disk. In each panel, red diamonds are calculated using the B13-based definition of M, blue squares represent the B95-based definition of M, and green diamonds represent the BD05-based definition of M. Panel (a) shows the C/M ratio vs. approximate [O/H] value at the center of each spatial bin. Panel (b) shows the C/M ratio vs approximate [O/H] value at the center of bins along the major axis only. Panel (c) shows the C/M ratio vs. mean photometric metallicity for all bins. Panel (d) shows the C/M ratio vs. mean photometric metallicity for bins along the major axis only.}
\label{fig:CMvs}
\end{center}
\end{figure*}

Panel (a) of Figure~\ref{fig:CMvs} show the resulting relationship between log(C/M) and [O/H]. Panel (b) shows the same relationship for only those bins along the major axis. There is a negative gradient, with the C/M ratio decreasing with increasing metallicity. We find the weighted best-fit to the major axis bins to be:
\begin{flalign*}
{\rm log(C/M_{B13})} &= (-0.96\! \pm  \!0.28)\!\times \!{\rm [O/H]}\! + (7.31\! \pm \!2.48)  &\\
{\rm log(C/M_{B95})} &=(-0.95\! \pm \!0.29)\!\times \!{\rm [O/H]}\! + \!(7.53\! \pm \!2.55)\\
{\rm log(C/M_{BD05})} &= (-0.90\! \pm \!0.29)\! \times \!{\rm [O/H]} \!+ \!(7.10\! \pm \!2.58)
\end{flalign*}

\renewcommand{\thefigure}{\arabic{figure} (Cont.)}
\addtocounter{figure}{-1}

\begin{figure*}
\begin{center}
\includegraphics[width = 6.5in]{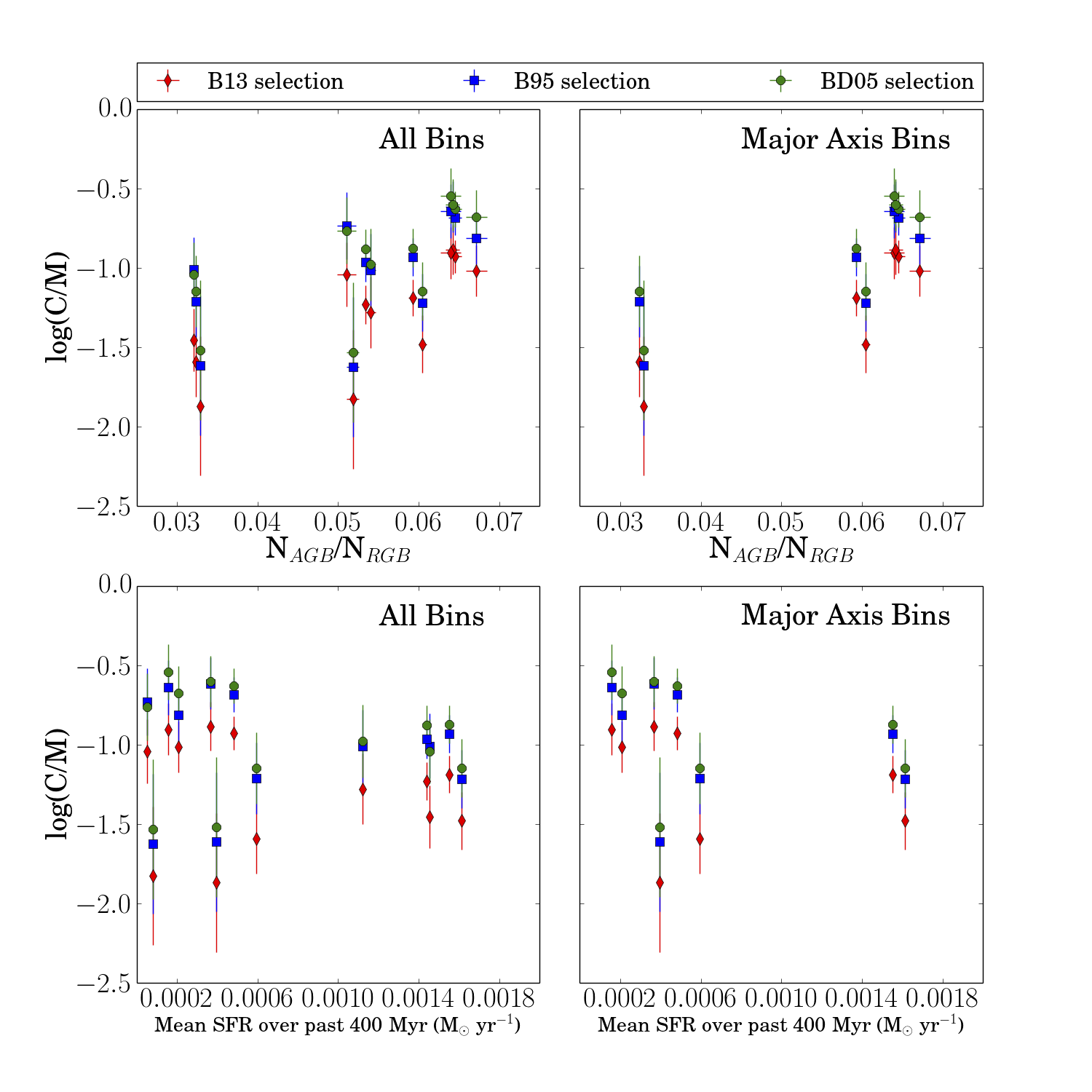}
\caption{Panel (e) shows the C/M ratio vs. our age proxy, N$_{AGB}$/N$_{RGB}$, for all spatial bins. Panel (f) shows the C/M ratio against bins along the major axis only. Panel (g) shows the C/M ratio vs. the mean SFR over the past 400~Myr for all spatial bins. Panel (h) shows the C/M ratio against the mean recent SFR for bins along the major axis only. }
\end{center}
\end{figure*}

\renewcommand{\thefigure}{\arabic{figure}}

H{\sc II} regions probe the metallicity of present-day star forming regions and young stars, not the relevant fraction of AGB stars that formed several Gyr ago. As a result, correlations between [O/H] in H{\sc II} regions and the C/M ratio tell us more about the metallicity evolution of the galaxy than the formation and evolution of the AGB stars themselves. An alternative source of gas-phase oxygen abundances are planetary nebulae (PNe), whose ages are much closer to those of AGB stars. 

\citet{Sanders2012} compute the oxygen abundance of PNe in the disk of M31. Unfortunately, there are not enough PNe in the vicinity of our data for us to look at the C/M ratio as a function of their [O/H] measurements (see Figure~\ref{fig:OHmap}). Recent work by \citet{Kwitter2012} and \citet{Balick2013} compute oxygen abundances for PNe in the outer disk of M31 ($R_{\rm g} >18$~kpc). Upcoming spectroscopic follow-up of the many PNe in the PHAT footprint \citep{Veyette2014} will provide reliable [O/H] measurements against which we can compare the C/M ratio.

\subsubsection{Stellar metallicity}\label{sec:Z}
In addition to a gas-phase metallicity, we can study the C/M ratio as a function of stellar metallicity. We use the photometric metallicity estimates derived for RGB stars in the PHAT fields by G15. While the photometric metallicity of the RGB population may not exactly match the metallicity of the AGB population, it is a far better tracer than H{\sc II} regions. In addition, we find that the combination of many more stars and more robust models makes photometric metallicity estimates derived from the RGB stars much more reliable than those we can derive using the spectra of our AGB stars.

G15 determine stellar metallicities by interpolating bias and completeness corrected photometry onto Padova PARSEC1.2s isochrones \citep{Bressan2012}, with a fixed age of 4~Gyr and a metallicity range $-2.18 < {\rm [M/H]} < 0.6$ ($0.0001 < {\rm Z} < 0.06$). To avoid being heavily biased by very crowded regions, they restrict their interpolation to those stars with $F814W < 23$. In the outer regions of the disk this cut corresponds to $100\%$ completeness, while in innermost regions it marks $\sim50\%$ completeness. Within the spatial bins shown in Figure~\ref{fig:map}, we compute the mean metallicity of all RGB stars satisfying the aforementioned magnitude and metallicity criteria. We leave out the metallicities of stars in PHAT Brick 1, where we have no spectra, so as not to artificially boost the average metallicity of the innermost bins beyond what is representative of our spectral data.

Figure~\ref{fig:CMvs}, panel (c), shows log(C/M) as a function of mean metallicity for all spatial bins. Panel (d) shows the relationship for only those bins along the major axis. We see a statistically significant negative gradient in C/M with photometric metallicity. A weighted best-fit to the spatial bins along the major axis returns
\begin{flalign*}
\rm log(C/M_{B13}) &= (-2.84 \! \pm \! 0.75)\! \times \! {\rm [M/H]}\! - \! (1.21 \! \pm \! 0.06) &\\ 
\rm log(C/M_{B95}) &=(-2.63 \! \pm \! 0.78) \! \times \! {\rm [M/H]}\! - \! (0.95 \! \pm \! 0.07)\\
\rm log(C/M_{BD05}) &= (-2.71 \! \pm \! 0.79)\! \times \! {\rm [M/H]}\! - \! (0.89 \! \pm \! 0.07)
\end{flalign*}

G15 note that the radial metallicity gradient computed from photometric RGB metallicities depends on the underlying age distribution, and provide median metallicity measurements for several fiducial ages. To test the dependence of the C/M ratio on age assumptions, we compute the gradient in log(C/M) with photometric metallicity for fiducial ages of 6~Gyr and 8~Gyr. For the B13 method, we find slopes of $-2.78 \pm 0.73$ and $-2.73 \pm 0.72$, respectively, which is perfectly consistent with the slope determined for a fiducial age of 4~Gyr. 

In addition to depending on the underlying age distribution, the metallicity determinations from G15 are affected by the choice of isochrones, low level dust extinction throughout the survey region, and any uncorrected photometric bias and completeness. As a result, the absolute metallicity measurements have large uncertainties.

\subsection{As a function of SFH}\label{sec:Age}

In addition to being a function of metallicity, the C/M ratio should depend on a galaxy's SFH. For a single-burst stellar population, the C/M ratio begins to increase from zero once turn-off masses reach 3-4~M$_\odot$. For ages older than a few Gyr, carbon stars are no longer produced and the C/M ratio falls again to zero. The exact age at which the C/M ratio peaks is determined by star formation rate (SFR) and metallicity \citep{Cioni2006, Marigo2013}. In a galaxy, this behavior is convolved with the SFH and smoothed. Using PHAT data, we can evaluate the C/M ratio as a function of population age and SFR.

While the exact ages of the AGB stars in M31 are difficult to compute, we can use the PHAT data to define a rough proxy. Again, we use the spatial bins shown in Figure~\ref{fig:map}. We define our proxy to be the number of TP-AGB stars in a given bin divided by the number of RGB stars (N$_{AGB}$/N$_{RGB}$). The value of this quantity as an age proxy relies on the fact that the mean progenitor mass of TP-AGB stars is larger, and hence younger, than that of RGB stars. A higher number ratio of TP-AGB stars to RGB stars implies a younger average population.

In order to be sensitive to the relative numbers of AGB and RGB stars independent of photometric completeness, we set a lower magnitude limit of $F814W = 23$.  We also leave out the data from PHAT Brick 1, so as not to be biased by the crowded bulge in which we have no spectra. With the crowded regions inward of $R_{\rm g} \sim 4$~kpc left out, our magnitude limit represents $100\%$ completeness in our bins. We also set a color limit of $F475W-F814W \ge 2$ to mitigate contamination by younger helium burning stars (see \S~\ref{phatID}).
\begin{figure}[t!]
\begin{center}
\includegraphics[width = 3.5in]{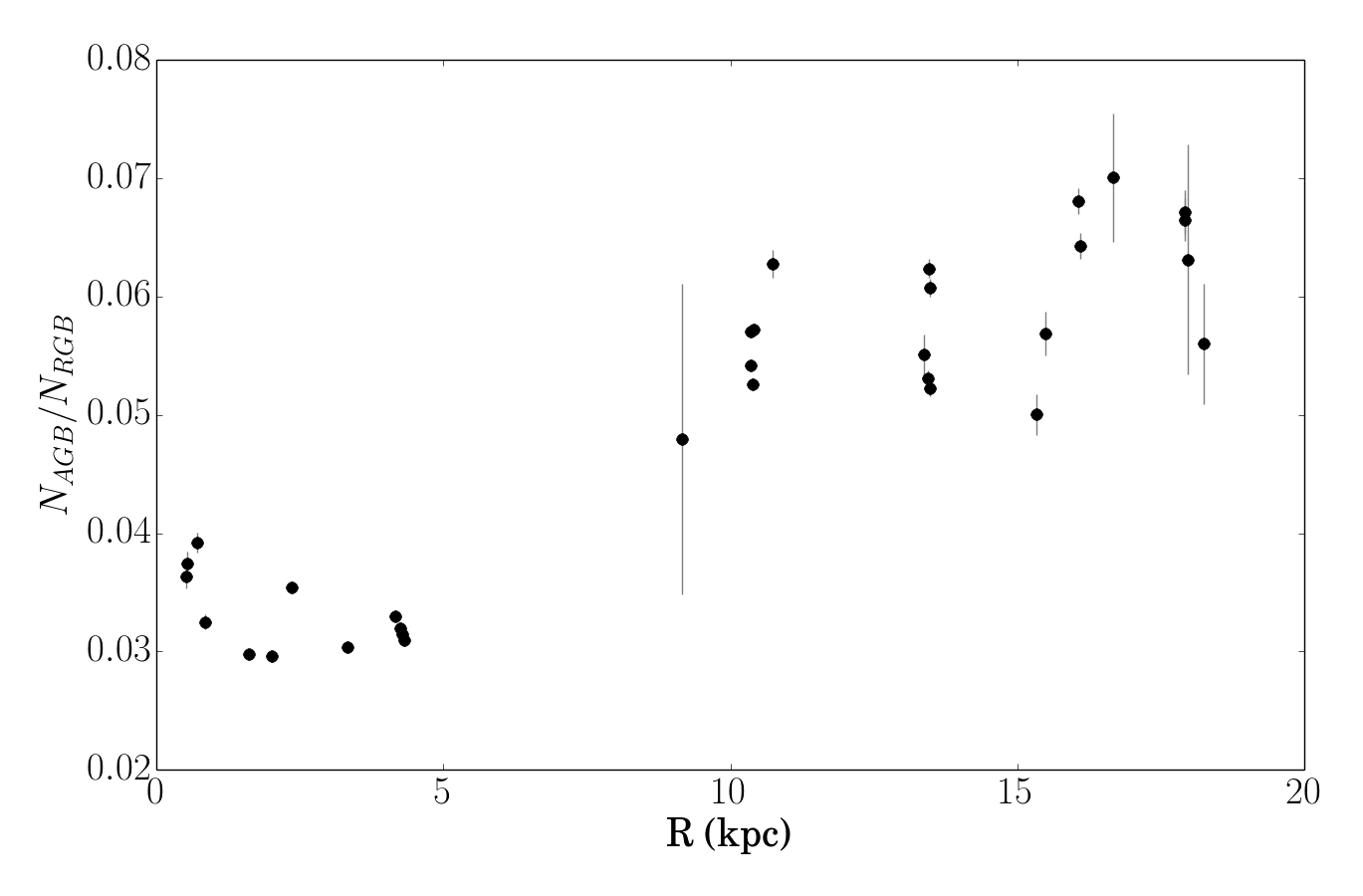}
\caption{Age proxy (N$_{AGB}$/N$_{RGB}$) as a function of $R_{\rm g}$ in kpc. Error bars represent $1\sigma$ Poisson uncertainties.}
\label{fig:age}
\end{center}
\end{figure}

A plot of our age proxy as a function of $R_{\rm g}$ is shown in Figure~\ref{fig:age}. As this verification does not require SPLASH data, whose smaller numbers limit how small we can bin, Figure~\ref{fig:age} is made with polar bins created by splitting M31 into 31 equal-area bins rather than the thirteen shown in Figure~\ref{fig:map}. As expected in the regime of hierarchical galaxy formation, we see that populations are younger as you move farther out in the disk. 

Figure~\ref{fig:CMvs}, panel (e), shows log(C/M) as a function of this age proxy computed in all spatial bins shown in Figure~\ref{fig:map} . Panel (f) shows the same for bins along the major axis. The weighted best-fit lines to bins along the major axis are:
\begin{flalign*}
\rm log(C/M_{\small B13}) &= (22.5 \! \pm \! 6.5)\! \times \! {N_{AGB}\over{N_{RGB}}}\! - \! (2.46 \! \pm \! 0.40) &\\ 
\rm log(C/M_{\small B95}) &= (18.8 \! \pm \! 6.6) \! \times \! {N_{AGB}\over{N_{RGB}}}\! - \! (1.98 \! \pm \! 0.40) \\
\rm log(C/M_{\small BD05}) &= (19.1 \! \pm \! 6.6)\! \times \! {N_{AGB}\over{N_{RGB}}}\! - \! (1.93 \! \pm \! 0.40) 
\end{flalign*}

We next look at the C/M ratio as a function of recent SFR, using data from \citet{Lewis2015} (hereafter L15). L15 apply the CMD fitting program MATCH \citep{Dolphin2002} to PHAT data in over 9000 regions in the disk of M31, and compute SFRs and cumulative stellar mass formation. To minimize the effect of model uncertainties their analysis focuses solely on the main sequence, and so probes the SFH of M31 over the last 400~Myr. While this timescale does not match the period during which the bulk of the AGB stars formed, both models and observations predict TP-AGB stars in populations as young as 100~Myr \citep[e.g.][]{Frogel1990}.
 
We take each of the 9000 regions investigated by L15 and group them by the spatial bins shown in Figure~\ref{fig:map} to compute a mean SFR. Log(C/M) as a function of mean recent SFR is shown in Figure~\ref{fig:CMvs} for all bins (panel g) and for bins along the major axis (panel h). There is considerably more scatter in these relationships than we saw for gas-phase oxygen abundance, stellar metallicity, or age proxy. However a weighted best-fit line to major axis bins indicates a statistically significant negative correlation. 
\begin{flalign*}
\rm log(C/M_{B13}) &= (-238 \! \pm \! 95)\! \times \! {\rm \overline{SFR}}\! - \! (0.91 \! \pm \! 0.09) &\\ 
\rm log(C/M_{B95}) &= (-229 \! \pm \! 98) \! \times \! {\rm \overline{SFR}}\! - \! (0.67 \! \pm \! 0.09) \\
\rm log(C/M_{BD05}) &= (-241 \! \pm \! 99)\! \times \! {\rm \overline{SFR}}- (0.59 \! \pm \! 0.09) 
\end{flalign*}
 
\section{Discussion}\label{discussion}

\subsection{The C/M ratio vs. environment}

\begin{deluxetable*}{lccccc}
\tablewidth{0pt}
\tablecaption{Coefficients of Multiple Regression Models \label{tab:regression}}
\tablehead{ & [O/H] &[M/H] & N$_{AGB}$/N$_{RGB}$ &SFR & R} \\

\startdata
Model 1 & & $-2.25 \pm 0.95$ & $9.61 \pm 7.7$ & & \\
Model 2 & $-0.59 \pm 0.24$ & $-0.99 \pm 0.86$ & $11.4 \pm 4.9$ & & \\
Model 3 & & $-1.79 \pm 1.48$ & $12.5 \pm 9.8$ & $-67.6 \pm 149$ & \\
Model 4 & & $-1.72 \pm 1.54$ & & & $0.04 \pm 0.03$ \\
Model 5 & $-0.64 \pm 0.17$ & & & & $0.05 \pm 0.01$ \\
Model 5 & & $-2.38 \pm 1.92$ & $10.6 \pm 16.2$ & & $-0.01 \pm 0.07$
\enddata
\tablecomments{Rows separate individual regression models, and columns represent which parameters were used in each model (gas-phase oxygen abundance [O/H], stellar metallicity [M/H], age N$_{AGB}$/N$_{RGB}$, mean SFR over the past 400~Myr, or galactocentric radius $R_{\rm g}$). A blank entry indicates that the parameter was not used in that particular regression model.}
\end{deluxetable*}

We find statistically significant trends between log(C/M) and each of the five properties we investigate: $R_{\rm g}$, present-day gas-phase [O/H], stellar [M/H], age proxy, and recent SFR. In each case, the gradient is made clearer by restricting the analysis to the eight spatial bins along the major axis. This is likely due to the increased uncertainty in the deprojection along the minor axis rather than an azimuthal change in the behavior of the C/M ratio.

The clearest trend we see is between log(C/M) and radius. Our slopes ($\sim 0.06$) are perfectly in line with the gradients observed by B95 (excluding their measurement at $\sim30$~kpc) and \citet{Battinelli2003} along the opposite axis of the M31 disk. While there is no reason the C/M ratio should depend on geometry, this correlation points to a significant relationship between the C/M ratio and properties that themselves vary with radius. 

B13 measure the C/M ratio in the inner disk of M31 ($R_{\rm g} = 2$~kpc), and find a C/M ratio much lower than predicted given the metallicity at that radius. They interpret this as an indication of a metallicity ceiling above which carbon stars do not form. If we extrapolate our radial gradient to $R_{\rm g} = 2$~kpc, we calculate log(C/M)$ = -1.81 \pm 0.22$. This is considerably higher than the value measured by B13 (log(C/M)$ = -3.48^{+0.85}_{-0.01}$). If the paucity of carbon stars observed in the inner disk by B13 was due to a gradual fall off in carbon star production with increasing metallicity, then our radial gradient should extrapolate to their measurement. The fact that the C/M ratio measured by B13 is so low is consistent with the suggestion that there is a hard metallicity limit above which carbon stars do not form.

While stellar oxygen abundance is the major driver of whether a TP-AGB star can become a carbon star, it is interesting that the [O/H] measurements from H{\sc II} regions in the disk of M31 show the expected trend when they themselves do not probe the metallicity of the AGB stars. One possible explanation for the observed gradient is that the metallicities of present-day star forming regions are correlated with the metallicities of star forming regions a few Gyr ago. This would indicate that the metallicity gradient in M31 is long lived. Alternatively, the C/M ratio and present-day gas-phase [O/H] may be independent, and the observed gradient is an artifact of their shared relationship with radius.  

The clear gradients we observe in log(C/M) with stellar metallicity and age are predicted by models and previous observations \citep[e.g.][]{Mouhcine2003, Feast2010, Held2010, Marigo2013}. Our observed relationship between log(C/M) and age indicates that young populations produce more M- than C-type TP-AGB stars, which could be explained by hot bottom burning operating on the most massive TP-AGB stars. The fact that the C/M ratio increases monotonically with our age proxy indicates that the average age of the populations in all our bins are older than a few Gyr. If this were not the case, then we would see the C/M ratio turn over and begin to decrease at the bins dominated by very young populations, where stars evolving off the main sequence do not become AGB stars or remain M-type due to hot bottom burning. Average ages consistently greater than a few Gyr is consistent with recent work showing a disk-wide burst of star formation 2-4~Gyr ago \citep{Williams2015}, and the age calculations by \citet{Dorman2015} and G15.

Our derived trend with metallicity is fully consistent with the relationship established across the Local Group \citep[log(C/M)$\sim (-2.12 \pm 0.04) \times {\rm [Fe/H]};$][]{Cioni2009}. The trend observed in Local Group satellites appears to extend smoothly to the metal-rich regime of M31, irrespective of their drastically different SFHs. This implies that the C/M ratio is strongly determined by the same population of intermediate-age stars in all environments, and depends only on stellar properties (metallicity and age). 

This interpretation of the log(C/M)-metallicity relationship would imply that our observed gradient in the C/M ratio with recent SFR is driven by correlations between SFR, age and metallicity rather than the impact of SFR itself on AGB evolution. The alternative explanation for that trend is that star formation in M31 is long lived, and the SFR of the past 400~Myr correlates with the SFR a few Gyr ago. This correlation could occur if the global burst of star formation 2-4~Gyr ago \citep{Williams2015} was spatially heterogeneous, and the regions of the disk producing the most stars 2-4~Gyr ago also have the highest recent SFR. 

To investigate whether we can decouple the effects of metallicity and age on the C/M ratio and determine the dominant factor, we construct a series of multiple regression models. Beginning with a simple linear relationship between log(C/M) and any property of the disk, analysis of variance finds that the model improvement from the addition of a second (or third) variable is insignificant. The coefficients of a sample of these models are shown in Table~\ref{tab:regression}; none are non-zero with high confidence. This is indicative of collinearity, and that age and metallicity are so tightly coupled that we cannot claim that statistically one is more important to the C/M ratio than the other. 

\subsection{Possible effects of dust}
Both ISM extinction within M31 and circumstellar dust may affect the measurement of the C/M ratio, and ISM extinction may affect our metallicity, age proxy and SFR measurements.

The C/M ratio will be affected by dust if it causes carbon stars or M-stars to be preferentially excluded from the SPLASH sample. Since we identify carbon stars spectroscopically, veiling of the prominent CN and C$_2$ features would make the stars unidentifiable. This is not the case for M-stars, as we identify them photometrically. However, recent models show that even the dustiest carbon stars do not have their $\sim7900$\AA~CN feature completely veiled (Aringer et.~al, private communication). It is far more likely that carbon stars escape our selection by being too dust-reddened to be visible in the optical, and are thus not targeted for spectroscopy.

To investigate the effects of interstellar dust, we use the M31 dust maps from \citet{Dalcanton2015} to determine the average extinction ($A_V$) at the position of each star. We compare the distributions of $A_V$ at the locations of the carbon stars and M-stars via a Kolmogorov-Smirnov (KS) test, and find no significant difference. Interstellar dust is thus not likely to have a significant effect on our selection. 

Circumstellar dust is known to have a major impact on the completeness of optical AGB surveys \citep[e.g.][]{Boyer2011a}. Correcting for AGB stars reddened out of the optical by circumstellar dust would steepen our measured age and metallicity gradients. In the case of metallicity, this is because while dust production in carbon stars stays constant as metallicity increases, dust production in M-stars goes up \citep{Sloan2008}. As metallicity increases the number of M-stars reddened from our sample will increase faster than the number of carbon stars reddened from our sample. The gradient with age proxy will steepen because carbon stars are typically more massive than M-stars, and produce more dust. Thus as the population gets younger, and has more massive stars climbing the AGB, the fraction of carbon stars invisible in the optical will be increasingly larger than the fraction of M-stars invisible in the optical. 

Because both age and metallicity decrease with radius in M31, both effects discussed above will contribute to our measuring a shallower gradient than reality.

By analyzing the $A_V$ and the fraction of reddened stars ($f_{red}$) G15 determine that their median metallicity measurements are generally not dependent on dust. They do find that the high metallicities found for stars along the north-west edge of the PHAT footprint track regions of higher dust extinction, however our spatial bins are large enough that these regions make up a small fraction of the stars do not substantially bias our median metallicities.

To evaluate the effect of dust on our age proxy, we first compare the distribution of extinction values in the vicinity of the AGB and RGB stars of the PHAT sample using the dust map from \citet{Dalcanton2015}. A KS-test indicates that we can reject the hypothesis that these populations are drawn from the same distribution of $A_V$. AGB stars are found in slightly dustier regions. However, we find no trend between age proxy and $A_V$ (R$^2 = 0.09$), which indicates that this does not heavily impact our results. 

L15 have taken dust into account when determining SFRs, so we will not further analyze the effect of dust on these measurements.

\subsection{Comparing definitions of M}

Throughout Section~\ref{results} we see that the definition of M in the C/M ratio has a significant impact on the intercept of gradients but a minimal impact on the slope. Consistently, the B95 and BD05 methods return very similar results, despite having rather different effective TRGBs. The B13 method, on the other hand, always counts more M giants and returns a lower C/M ratio. This result indicates that care must be taken when comparing values within the literature, as inhomogeneous methods lead to different absolute values of log(C/M). The fact that slopes are consistent between different selection methods while intercepts are not tells us that the population of stars not being included -- the bluer ``AGB" stars -- are distributed evenly across the disk. As we see gradients in log(C/M) with both age proxy and metallicity, this blue population is likely not the youngest or most metal poor AGB stars, but is instead either foreground contamination or supergiants.

The major result from B13, that the C/M ratio calculated $\sim 2$~kpc from the center of M31 is highly discrepant with values calculated farther out in the disk, stems from a comparison between the C/M ratios calculated by the B13 and B95 methods. B13 thoroughly investigate whether this discrepancy stems from an underestimation of the number of C-stars, but do not consider whether they are overestimating the number of M-stars (relative to the studies to which they compare). We can estimate how much the B13 data point would change if the B95 criteria were applied. In our data, using the B13 definition rather than the B95 definition amounts to a difference of $\sim800$ M stars, or a difference of $\sim0.35$ in log(C/M). While this moves the B13 point closer to the trend fit to B95 data, it is still significantly discrepant (off by $\sim1$ in log(C/M)). However, this rough approximation does not take into account the fact that B13 use medium-band NIR photometry to count M-stars while we use optical photometry. We cannot reproduce the effectiveness of medium-band NIR photometry, which is centered on distinctive absorption features present in either C or M stars, with PHAT wide-band NIR photometry. 

\section{Conclusion}

We have computed the C/M ratio in the disk of M31 using spectroscopic and photometric data from the SPLASH and PHAT surveys. An uncontaminated sample of carbon stars was identified using moderate-resolution optical spectra. M-stars were identified photometrically in three different ways, following methods used by \citet{Boyer2013}, \citet{Brewer1995} and \citet{BattinelliDemers2005}. We have calculated the C/M ratio as a function of galactocentric radius, present-day gas-phase metallicity, stellar metallicity, age (via proxy N$_{AGB}$/N$_{RGB}$), and mean SFR over the past 400~Myr. 

From this, we conclude:
\begin{itemize}

\item The definition of ``M" has a minimal effect on the slope of a relationship, but a substantial effect on the fiducial value. This adds to the body of evidence that stresses the need for homogenous samples when studying the C/M ratio as a function of other parameters.

\item There are statistically significant correlations between log(C/M) and $R_{\rm g}$, present-day gas-phase [O/H], stellar [M/H], age proxy, and recent SFR. 

\item We reproduce the relationship between log(C/M) and stellar metallicity stated in \citet{Cioni2009}, despite working in a substantially more metal-rich environment with a drastically different SFH. 

\item Age and metallicity are too closely connected to state that one is more important to the C/M ratio in M31 than the other. 

\end{itemize}
\acknowledgements

PG and KH acknowledge NSF grants AST-1010039 and AST-1412648. This work was also supported by NASA grant HST-GO-12055. KH was supported by a NSF Graduate Research Fellowship. DRW is supported by NASA through Hubble Fellowship grant HST-HF-51331.01 awarded by the Space Telescope Science Institute. We thank Bernhard Aringer, Zachary Jennings and Alex Rudy for helpful conversations. We appreciate the very significant cultural role and reverence that the summit of Mauna Kea has always had within the indigenous Hawaiian community. We are most grateful to have had the opportunity to conduct observations from this mountain.

\bibliographystyle{apj}
\bibliography{CtoM}

\begin{thebibliography}{64}
\expandafter\ifx\csname natexlab\endcsname\relax\def\natexlab#1{#1}\fi

\bibitem[{{Balick} {et~al.}(2013){Balick}, {Kwitter}, {Corradi}, \&
  {Henry}}]{Balick2013}
{Balick}, B., {Kwitter}, K.~B., {Corradi}, R.~L.~M., \& {Henry}, R.~B.~C. 2013,
  \apj, 774, 3

\bibitem[{{Barmby} {et~al.}(2006){Barmby}, {Ashby}, {Bianchi}, {Engelbracht},
  {Gehrz}, {Gordon}, {Hinz}, {Huchra}, {Humphreys}, {Pahre},
  {P{\'e}rez-Gonz{\'a}lez}, {Polomski}, {Rieke}, {Thilker}, {Willner}, \&
  {Woodward}}]{Barmby2006}
{Barmby}, P., {Ashby}, M.~L.~N., {Bianchi}, L., {Engelbracht}, C.~W., {Gehrz},
  R.~D., {Gordon}, K.~D., {Hinz}, J.~L., {Huchra}, J.~P., {Humphreys}, R.~M.,
  {Pahre}, M.~A., {P{\'e}rez-Gonz{\'a}lez}, P.~G., {Polomski}, E.~F., {Rieke},
  G.~H., {Thilker}, D.~A., {Willner}, S.~P., \& {Woodward}, C.~E. 2006, \apjl,
  650, L45

\bibitem[{Battinelli \& Demers(2004{\natexlab{a}})}]{Battinelli2004a}
Battinelli, P. \& Demers, S. 2004{\natexlab{a}}, Astronomy \& Astrophysics,
  418, 33

\bibitem[{Battinelli \& Demers(2004{\natexlab{b}})}]{BattinelliDemers2004}
---. 2004{\natexlab{b}}, Astronomy \& Astrophysics, 417, 479

\bibitem[{Battinelli \& Demers(2004{\natexlab{c}})}]{Battinelli2004}
---. 2004{\natexlab{c}}, Astronomy \& Astrophysics, 417, 479

\bibitem[{Battinelli \& Demers(2005)}]{BattinelliDemers2005}
---. 2005, Astronomy \& Astrophysics, 430, 905

\bibitem[{{Battinelli} \& {Demers}(2005)}]{Battinelli2005}
{Battinelli}, P. \& {Demers}, S. 2005, \aap, 434, 657

\bibitem[{Battinelli \& Demers(2009{\natexlab{a}})}]{BattinelliDemers2009}
Battinelli, P. \& Demers, S. 2009{\natexlab{a}}, Astronomy \& Astrophysics,
  493, 1075

\bibitem[{Battinelli \& Demers(2009{\natexlab{b}})}]{Battinelli2009}
---. 2009{\natexlab{b}}, Astronomy \& Astrophysics, 493, 1075

\bibitem[{Battinelli {et~al.}(2003)Battinelli, Demers, \&
  Letarte}]{Battinelli2003}
Battinelli, P., Demers, S., \& Letarte, B. 2003, The Astronomical Journal, 125,
  1298

\bibitem[{Boyer {et~al.}(2013)Boyer, Girardi, Marigo, Williams, Aringer,
  Nowotny, Rosenfield, Dorman, Guhathakurta, Dalcanton, Melbourne, Olsen, \&
  Weisz}]{Boyer2013}
Boyer, M.~L., Girardi, L., Marigo, P., Williams, B.~F., Aringer, B., Nowotny,
  W., Rosenfield, P., Dorman, C.~E., Guhathakurta, P., Dalcanton, J.~J.,
  Melbourne, J.~L., Olsen, K. a.~G., \& Weisz, D.~R. 2013, The Astrophysical
  Journal, 774, 83

\bibitem[{Boyer {et~al.}(2011)Boyer, Srinivasan, van Loon, McDonald, Meixner,
  Zaritsky, Gordon, Kemper, Babler, Block, Bracker, Engelbracht, Hora,
  Indebetouw, Meade, Misselt, Robitaille, Sewiło, Shiao, Whitney, \&
  Sewił}]{Boyer2011a}
Boyer, M.~L., Srinivasan, S., van Loon, J.~T., McDonald, I., Meixner, M.,
  Zaritsky, D., Gordon, K.~D., Kemper, F., Babler, B., Block, M., Bracker, S.,
  Engelbracht, C.~W., Hora, J., Indebetouw, R., Meade, M., Misselt, K.,
  Robitaille, T., Sewiło, M., Shiao, B., Whitney, B., \& Sewił, M. 2011, The
  Astronomical Journal, 142, 103

\bibitem[{Bressan {et~al.}(2012)Bressan, Marigo, Girardi, Salasnich, {Dal
  Cero}, Rubele, \& Nanni}]{Bressan2012}
Bressan, A., Marigo, P., Girardi, L., Salasnich, B., {Dal Cero}, C., Rubele,
  S., \& Nanni, A. 2012, Monthly Notices of the Royal Astronomical Society,
  427, 127

\bibitem[{Brewer {et~al.}(1995)Brewer, Richer, \& Crabtree}]{Brewer1995}
Brewer, J.~P., Richer, H.~B., \& Crabtree, D.~R. 1995, The Astronomical
  Journal, 109

\bibitem[{Brewer {et~al.}(1996)Brewer, Richer, \& Crabtree}]{Brewer1996}
---. 1996, The Astronomical Journal, 112

\bibitem[{{Chen} {et~al.}(2014){Chen}, {Trager}, {Peletier}, {Lan{\c c}on},
  {Vazdekis}, {Prugniel}, {Silva}, {Gonneau}, {Lyubenova}, {Koleva}, {Barroso},
  {Bl{\'a}zquez}, {Walcher}, {Choudhury}, \& {Meneses-Goytia}}]{Chen2014}
{Chen}, Y.-P., {Trager}, S.~C., {Peletier}, R.~F., {Lan{\c c}on}, A.,
  {Vazdekis}, A., {Prugniel}, P., {Silva}, D., {Gonneau}, A., {Lyubenova}, M.,
  {Koleva}, M., {Barroso}, J.~F., {Bl{\'a}zquez}, P.~S., {Walcher}, C.~J.,
  {Choudhury}, O.~S., \& {Meneses-Goytia}, S. 2014, The Messenger, 158, 30

\bibitem[{Cioni(2009)}]{Cioni2009}
Cioni, M.-R.~L. 2009, Astronomy and Astrophysics, 506, 1137

\bibitem[{{Cioni} {et~al.}(2006){Cioni}, {Girardi}, {Marigo}, \&
  {Habing}}]{Cioni2006}
{Cioni}, M.-R.~L., {Girardi}, L., {Marigo}, P., \& {Habing}, H.~J. 2006, \aap,
  448, 77

\bibitem[{{Cioni} \& {Habing}(2003)}]{Cioni2003}
{Cioni}, M.-R.~L. \& {Habing}, H.~J. 2003, \aap, 402, 133

\bibitem[{{Cioni} {et~al.}(2008){Cioni}, {Irwin}, {Ferguson}, {McConnachie},
  {Conn}, {Huxor}, {Ibata}, {Lewis}, \& {Tanvir}}]{Cioni2008}
{Cioni}, M.-R.~L., {Irwin}, M., {Ferguson}, A.~M.~N., {McConnachie}, A.,
  {Conn}, B.~C., {Huxor}, A., {Ibata}, R., {Lewis}, G., \& {Tanvir}, N. 2008,
  \aap, 487, 131

\bibitem[{{Conroy}(2013)}]{Conroy2013}
{Conroy}, C. 2013, \araa, 51, 393

\bibitem[{{Cooper} {et~al.}(2012){Cooper}, {Newman}, {Davis}, {Finkbeiner}, \&
  {Gerke}}]{Cooper2012}
{Cooper}, M.~C., {Newman}, J.~A., {Davis}, M., {Finkbeiner}, D.~P., \& {Gerke},
  B.~F. 2012, {spec2d: DEEP2 DEIMOS Spectral Pipeline}, Astrophysics Source
  Code Library

\bibitem[{Dalcanton {et~al.}(2015)Dalcanton, Fouesneau, W., Lang, Leroy,
  Gordon, Sandstrom, Weisz, Williams, Bell, Dong, Gilber, Girardi, Gouliermis,
  Lauer, Seth, Schruba, \& Skillman}]{Dalcanton2015}
Dalcanton, J.~J., Fouesneau, M., W., H.~D., Lang, D., Leroy, A., Gordon, K.~G.,
  Sandstrom, K., Weisz, D.~R., Williams, B.~F., Bell, E.~F., Dong, H., Gilber,
  K.~M., Girardi, L., Gouliermis, D.~A., Lauer, T., Seth, A., Schruba, A., \&
  Skillman, E.~D. 2015, \apj

\bibitem[{Dalcanton {et~al.}(2012)Dalcanton, Williams, Lang, Lauer, Kalirai,
  Seth, Dolphin, Rosenfield, Weisz, Bell, Bianchi, Boyer, Caldwell, Dong,
  Dorman, Gilbert, Girardi, Gogarten, Gordon, Guhathakurta, Hodge, Holtzman,
  Johnson, Larsen, Lewis, Melbourne, Olsen, Rix, Rosema, Saha, Sarajedini,
  Skillman, \& Stanek}]{Dalcanton2012}
Dalcanton, J.~J., Williams, B.~F., Lang, D., Lauer, T.~R., Kalirai, J.~S.,
  Seth, A.~C., Dolphin, A., Rosenfield, P., Weisz, D.~R., Bell, E.~F., Bianchi,
  L.~C., Boyer, M.~L., Caldwell, N., Dong, H., Dorman, C.~E., Gilbert, K.~M.,
  Girardi, L., Gogarten, S.~M., Gordon, K.~D., Guhathakurta, P., Hodge, P.~W.,
  Holtzman, J.~a., Johnson, L.~C., Larsen, S. r.~S., Lewis, A., Melbourne,
  J.~L., Olsen, K. a.~G., Rix, H.-W., Rosema, K., Saha, A., Sarajedini, A.,
  Skillman, E.~D., \& Stanek, K.~Z. 2012, The Astrophysical Journal Supplement
  Series, 200

\bibitem[{{Davidge}(2001)}]{Davidge2001}
{Davidge}, T.~J. 2001, \aj, 122, 1386

\bibitem[{{Davidge} {et~al.}(2005){Davidge}, {Olsen}, {Blum}, {Stephens}, \&
  {Rigaut}}]{Davidge2005}
{Davidge}, T.~J., {Olsen}, K.~A.~G., {Blum}, R., {Stephens}, A.~W., \&
  {Rigaut}, F. 2005, \aj, 129, 201

\bibitem[{Dolphin(2002)}]{Dolphin2002}
Dolphin, A.~E. 2002, Monthly Notices of the Royal Astronomical Society, 332, 91

\bibitem[{Dorman {et~al.}(2012)Dorman, Guhathakurta, Fardal, Lang, Geha,
  Howley, Kalirai, Bullock, Cuillandre, Dalcanton, Gilbert, Seth, Tollerud,
  Williams, \& Yniguez}]{Dorman2012}
Dorman, C.~E., Guhathakurta, P., Fardal, M.~a., Lang, D., Geha, M.~C., Howley,
  K.~M., Kalirai, J.~S., Bullock, J.~S., Cuillandre, J.-C., Dalcanton, J.~J.,
  Gilbert, K.~M., Seth, A.~C., Tollerud, E.~J., Williams, B.~F., \& Yniguez, B.
  2012, The Astrophysical Journal, 752, 147

\bibitem[{{Dorman} {et~al.}(2015){Dorman}, {Guhathakurta}, {Seth}, {Weisz},
  {Bell}, {Dalcanton}, {Gilbert}, {Hamren}, {Lewis}, {Skillman}, {Toloba}, \&
  {Williams}}]{Dorman2015}
{Dorman}, C.~E., {Guhathakurta}, P., {Seth}, A.~C., {Weisz}, D.~R., {Bell},
  E.~F., {Dalcanton}, J.~J., {Gilbert}, K.~M., {Hamren}, K.~M., {Lewis}, A.~R.,
  {Skillman}, E.~D., {Toloba}, E., \& {Williams}, B.~F. 2015, ArXiv e-prints

\bibitem[{{Faber} {et~al.}(2003){Faber}, {Phillips}, {Kibrick}, {Alcott},
  {Allen}, {Burrous}, {Cantrall}, {Clarke}, {Coil}, {Cowley}, {Davis}, {Deich},
  {Dietsch}, {Gilmore}, {Harper}, {Hilyard}, {Lewis}, {McVeigh}, {Newman},
  {Osborne}, {Schiavon}, {Stover}, {Tucker}, {Wallace}, {Wei}, {Wirth}, \&
  {Wright}}]{Faber2003}
{Faber}, S.~M., {Phillips}, A.~C., {Kibrick}, R.~I., {Alcott}, B., {Allen},
  S.~L., {Burrous}, J., {Cantrall}, T., {Clarke}, D., {Coil}, A.~L., {Cowley},
  D.~J., {Davis}, M., {Deich}, W.~T.~S., {Dietsch}, K., {Gilmore}, D.~K.,
  {Harper}, C.~A., {Hilyard}, D.~F., {Lewis}, J.~P., {McVeigh}, M., {Newman},
  J., {Osborne}, J., {Schiavon}, R., {Stover}, R.~J., {Tucker}, D., {Wallace},
  V., {Wei}, M., {Wirth}, G., \& {Wright}, C.~A. 2003, in Society of
  Photo-Optical Instrumentation Engineers (SPIE) Conference Series, Vol. 4841,
  Instrument Design and Performance for Optical/Infrared Ground-based
  Telescopes, ed. M.~{Iye} \& A.~F.~M. {Moorwood}, 1657--1669

\bibitem[{{Feast} {et~al.}(2010){Feast}, {Abedigamba}, \&
  {Whitelock}}]{Feast2010}
{Feast}, M.~W., {Abedigamba}, O.~P., \& {Whitelock}, P.~A. 2010, \mnras, 408,
  L76

\bibitem[{Fluks {et~al.}(1994)Fluks, Plez, The, de~Winter, Westerlund, \&
  Steenman}]{Fluks1994}
Fluks, M.~A., Plez, B., The, P.~S., de~Winter, D., Westerlund, B., \& Steenman,
  H.~C. 1994, Astronomy and Astrophysics Supplement Series, 105, 311

\bibitem[{Freedman \& Madore(1990)}]{Freedman1990}
Freedman, W. \& Madore, B.~F. 1990, The Astrophysical Journal, 365, 186

\bibitem[{{Frogel} {et~al.}(1990){Frogel}, {Mould}, \& {Blanco}}]{Frogel1990}
{Frogel}, J.~A., {Mould}, J., \& {Blanco}, V.~M. 1990, \apj, 352, 96

\bibitem[{Gilbert {et~al.}(2009)Gilbert, Font, Johnston, \&
  Guhathakurta}]{Gilbert2009}
Gilbert, K.~M., Font, A.~S., Johnston, K.~V., \& Guhathakurta, P. 2009, The
  Astrophysical Journal, 701, 776

\bibitem[{Gilbert {et~al.}(2012)Gilbert, Guhathakurta, Beaton, Bullock, Geha,
  Kalirai, Kirby, Majewski, Ostheimer, Patterson, Tollerud, Tanaka, \&
  Chiba}]{Gilbert2012}
Gilbert, K.~M., Guhathakurta, P., Beaton, R.~L., Bullock, J., Geha, M.~C.,
  Kalirai, J.~S., Kirby, E.~N., Majewski, S.~R., Ostheimer, J.~C., Patterson,
  R.~J., Tollerud, E.~J., Tanaka, M., \& Chiba, M. 2012, The Astrophysical
  Journal, 760, 76

\bibitem[{{Gregersen} {et~al.}(2015){Gregersen}, {Seth}, {Williams}, {Lang},
  {Dalcanton}, {Girardi}, {Skillman}, {Bell}, {Dolpin}, {Fouseneau}, {Hamren},
  {Johnson}, {Kalirai}, {Lewis}, {Monachesi}, \& {Olsen}}]{Gregersen2015}
{Gregersen}, D., {Seth}, A., {Williams}, B.~F., {Lang}, D., {Dalcanton}, J.~J.,
  {Girardi}, L., {Skillman}, E.~D., {Bell}, E., {Dolpin}, A.~E., {Fouseneau},
  M., {Hamren}, K., {Johnson}, L.~C., {Kalirai}, J., {Lewis}, A., {Monachesi},
  A., \& {Olsen}, K. 2015, submitted

\bibitem[{Guhathakurta {et~al.}(2006)Guhathakurta, Rich, Reitzel, Cooper,
  Gilbert, Majewski, Ostheimer, Geha, Johnston, \&
  Patterson}]{Guhathakurta2006a}
Guhathakurta, P., Rich, R.~M., Reitzel, D.~B., Cooper, M.~C., Gilbert, K.~M.,
  Majewski, S.~R., Ostheimer, J.~C., Geha, M.~C., Johnston, K.~V., \&
  Patterson, R.~J. 2006, The Astronomical Journal, 131, 2497

\bibitem[{{Held} {et~al.}(2010){Held}, {Gullieuszik}, {Rizzi}, {Girardi},
  {Marigo}, \& {Saviane}}]{Held2010}
{Held}, E.~V., {Gullieuszik}, M., {Rizzi}, L., {Girardi}, L., {Marigo}, P., \&
  {Saviane}, I. 2010, \mnras, 404, 1475

\bibitem[{Karakas(2014)}]{Karakas2014}
Karakas, A. 2014, Monthly Notices of the Royal Astronomical Society, 000

\bibitem[{{Keenan} \& {Morgan}(1941)}]{KeenanMorgan1941}
{Keenan}, P.~C. \& {Morgan}, W.~W. 1941, \apj, 94, 501

\bibitem[{Kewley \& Dopita(2002)}]{Kewley2002}
Kewley, L.~J. \& Dopita, M.~A. 2002, The Astrophysical Journal Supplement
  Series, 142, 35

\bibitem[{{Kwitter} {et~al.}(2012){Kwitter}, {Lehman}, {Balick}, \&
  {Henry}}]{Kwitter2012}
{Kwitter}, K.~B., {Lehman}, E.~M.~M., {Balick}, B., \& {Henry}, R.~B.~C. 2012,
  \apj, 753, 12

\bibitem[{{Lewis} {et~al.}(2015){Lewis}, {Dolphin}, {Dalcanton}, {Weisz},
  {Williams}, {Bell}, {Seth}, {Simones}, {Skillman}, {Choi}, {Fouesneau},
  {Guhathakurta}, {Johnson}, {Kalirai}, {Leroy}, {Monachesi}, {Rix}, \&
  {Schruba}}]{Lewis2015}
{Lewis}, A.~R., {Dolphin}, A.~E., {Dalcanton}, J.~J., {Weisz}, D.~R.,
  {Williams}, B.~F., {Bell}, E.~F., {Seth}, A.~C., {Simones}, J.~E.,
  {Skillman}, E.~D., {Choi}, Y., {Fouesneau}, M., {Guhathakurta}, P.,
  {Johnson}, L.~C., {Kalirai}, J.~S., {Leroy}, A.~K., {Monachesi}, A., {Rix},
  H.-W., \& {Schruba}, A. 2015, ArXiv e-prints

\bibitem[{{Marigo} {et~al.}(2013){Marigo}, {Bressan}, {Nanni}, {Girardi}, \&
  {Pumo}}]{Marigo2013}
{Marigo}, P., {Bressan}, A., {Nanni}, A., {Girardi}, L., \& {Pumo}, M.~L. 2013,
  \mnras, 434, 488

\bibitem[{Melbourne \& Boyer(2013)}]{Melbourne2013}
Melbourne, J. \& Boyer, M.~L. 2013, The Astrophysical Journal, 764, 30

\bibitem[{Melbourne {et~al.}(2012)Melbourne, Williams, Dalcanton, Rosenfield,
  Girardi, Marigo, Weisz, Dolphin, Boyer, Olsen, Skillman, \&
  Seth}]{Melbourne2012}
Melbourne, J., Williams, B.~F., Dalcanton, J.~J., Rosenfield, P., Girardi, L.,
  Marigo, P., Weisz, D., Dolphin, A., Boyer, M.~L., Olsen, K., Skillman, E., \&
  Seth, A.~C. 2012, The Astrophysical Journal, 748, 47

\bibitem[{{Menzies} {et~al.}(2015){Menzies}, {Whitelock}, \&
  {Feast}}]{Menzies2015}
{Menzies}, J.~W., {Whitelock}, P.~A., \& {Feast}, M.~W. 2015, ArXiv e-prints

\bibitem[{{Mouhcine} \& {Lan{\c c}on}(2003)}]{Mouhcine2003}
{Mouhcine}, M. \& {Lan{\c c}on}, A. 2003, \mnras, 338, 572

\bibitem[{Nagao {et~al.}(2006)Nagao, Maiolino, \& Marconi}]{Nagao2006}
Nagao, T., Maiolino, R., \& Marconi, A. 2006, Astronomy \& Astrophysics, 459,
  85

\bibitem[{{Newman} {et~al.}(2013){Newman}, {Cooper}, {Davis}, {Faber}, {Coil},
  {Guhathakurta}, {Koo}, {Phillips}, {Conroy}, {Dutton}, {Finkbeiner}, {Gerke},
  {Rosario}, {Weiner}, {Willmer}, {Yan}, {Harker}, {Kassin}, {Konidaris},
  {Lai}, {Madgwick}, {Noeske}, {Wirth}, {Connolly}, {Kaiser}, {Kirby},
  {Lemaux}, {Lin}, {Lotz}, {Luppino}, {Marinoni}, {Matthews}, {Metevier}, \&
  {Schiavon}}]{Newman2013}
{Newman}, J.~A., {Cooper}, M.~C., {Davis}, M., {Faber}, S.~M., {Coil}, A.~L.,
  {Guhathakurta}, P., {Koo}, D.~C., {Phillips}, A.~C., {Conroy}, C., {Dutton},
  A.~A., {Finkbeiner}, D.~P., {Gerke}, B.~F., {Rosario}, D.~J., {Weiner},
  B.~J., {Willmer}, C.~N.~A., {Yan}, R., {Harker}, J.~J., {Kassin}, S.~A.,
  {Konidaris}, N.~P., {Lai}, K., {Madgwick}, D.~S., {Noeske}, K.~G., {Wirth},
  G.~D., {Connolly}, A.~J., {Kaiser}, N., {Kirby}, E.~N., {Lemaux}, B.~C.,
  {Lin}, L., {Lotz}, J.~M., {Luppino}, G.~A., {Marinoni}, C., {Matthews},
  D.~J., {Metevier}, A., \& {Schiavon}, R.~P. 2013, \apjs, 208, 5

\bibitem[{Nowotny {et~al.}(2003)Nowotny, Kerschbaum, Olofsson, \&
  Schwarz}]{Nowotny2003}
Nowotny, W., Kerschbaum, F., Olofsson, H., \& Schwarz, H.~E. 2003, Astronomy \&
  Astrophysics, 403, 93

\bibitem[{Pilyugin \& Thuan(2005)}]{Pilyugin2005}
Pilyugin, L.~S. \& Thuan, T.~X. 2005, The Astrophysical Journal, 631, 231

\bibitem[{Sanders {et~al.}(2012)Sanders, Caldwell, McDowell, \&
  Harding}]{Sanders2012}
Sanders, N.~E., Caldwell, N., McDowell, J., \& Harding, P. 2012, The
  Astrophysical Journal, 758, 133

\bibitem[{Sirianni {et~al.}(2005)Sirianni, Jee, Ben\'{\i}tez, Blakeslee,
  Martel, Meurer, Clampin, Marchi, Ford, Gilliland, Hartig, Illingworth, Mack,
  Mccann, Sirianni, \& Benı}]{Sirianni2005}
Sirianni, A.~M., Jee, M.~J., Ben\'{\i}tez, N., Blakeslee, J.~P., Martel, A.~R.,
  Meurer, G., Clampin, M., Marchi, G.~D., Ford, H.~C., Gilliland, R., Hartig,
  G.~F., Illingworth, G.~D., Mack, J., Mccann, W.~J., Sirianni, M., \& Benı,
  N. 2005, Publications of the Astronomical Society of the Pacific, 117, 1049

\bibitem[{{Sloan} {et~al.}(2008){Sloan}, {Kraemer}, {Wood}, {Zijlstra},
  {Bernard-Salas}, {Devost}, \& {Houck}}]{Sloan2008}
{Sloan}, G.~C., {Kraemer}, K.~E., {Wood}, P.~R., {Zijlstra}, A.~A.,
  {Bernard-Salas}, J., {Devost}, D., \& {Houck}, J.~R. 2008, \apj, 686, 1056

\bibitem[{Stephens {et~al.}(2003)Stephens, Frogel, DePoy, Freedman, Gallart,
  Jablonka, Renzini, Rich, \& Davies}]{Stephens2003}
Stephens, A.~W., Frogel, J.~A., DePoy, D.~L., Freedman, W., Gallart, C.,
  Jablonka, P., Renzini, A., Rich, R.~M., \& Davies, R. 2003, The Astronomical
  Journal, 125, 2473

\bibitem[{{Veyette} {et~al.}(2014){Veyette}, {Williams}, {Dalcanton}, {Balick},
  {Caldwell}, {Fouesneau}, {Girardi}, {Gordon}, {Kalirai}, {Rosenfield}, \&
  {Seth}}]{Veyette2014}
{Veyette}, M.~J., {Williams}, B.~F., {Dalcanton}, J.~J., {Balick}, B.,
  {Caldwell}, N., {Fouesneau}, M., {Girardi}, L., {Gordon}, K.~D., {Kalirai},
  J., {Rosenfield}, P., \& {Seth}, A.~C. 2014, \apj, 792, 121

\bibitem[{{Villaume} {et~al.}(2015){Villaume}, {Conroy}, \&
  {Johnson}}]{Villaume2015}
{Villaume}, A., {Conroy}, C., \& {Johnson}, B. 2015, ArXiv e-prints

\bibitem[{{Wenger} {et~al.}(2000){Wenger}, {Ochsenbein}, {Egret}, {Dubois},
  {Bonnarel}, {Borde}, {Genova}, {Jasniewicz}, {Lalo{\"e}}, {Lesteven}, \&
  {Monier}}]{Wenger2000}
{Wenger}, M., {Ochsenbein}, F., {Egret}, D., {Dubois}, P., {Bonnarel}, F.,
  {Borde}, S., {Genova}, F., {Jasniewicz}, G., {Lalo{\"e}}, S., {Lesteven}, S.,
  \& {Monier}, R. 2000, \aaps, 143, 9

\bibitem[{{Williams} {et~al.}(2015){Williams}, {Dalcanton}, {Dolphin}, {Weisz},
  {Lewis}, {Lang}, {Bell}, {Boyer}, {Fouesneau}, {Gilbert}, {Monachesi}, \&
  {Skillman}}]{Williams2015}
{Williams}, B.~F., {Dalcanton}, J.~J., {Dolphin}, A.~E., {Weisz}, D.~R.,
  {Lewis}, A.~R., {Lang}, D., {Bell}, E.~F., {Boyer}, M., {Fouesneau}, M.,
  {Gilbert}, K.~M., {Monachesi}, A., \& {Skillman}, E. 2015, ArXiv e-prints

\bibitem[{{Williams} {et~al.}(2014){Williams}, {Lang}, {Dalcanton}, {Dolphin},
  {Weisz}, {Bell}, {Bianchi}, {Byler}, {Gilbert}, {Girardi}, {Gordon},
  {Gregersen}, {Johnson}, {Kalirai}, {Lauer}, {Monachesi}, {Rosenfield},
  {Seth}, \& {Skillman}}]{Williams2014}
{Williams}, B.~F., {Lang}, D., {Dalcanton}, J.~J., {Dolphin}, A.~E., {Weisz},
  D.~R., {Bell}, E.~F., {Bianchi}, L., {Byler}, N., {Gilbert}, K.~M.,
  {Girardi}, L., {Gordon}, K., {Gregersen}, D., {Johnson}, L.~C., {Kalirai},
  J., {Lauer}, T.~R., {Monachesi}, A., {Rosenfield}, P., {Seth}, A., \&
  {Skillman}, E. 2014, \apjs, 215, 9

\bibitem[{{Wing}(2007)}]{Wing2007}
{Wing}, R.~F. 2007, in Astronomical Society of the Pacific Conference Series,
  Vol. 378, Why Galaxies Care About AGB Stars: Their Importance as Actors and
  Probes, ed. F.~{Kerschbaum}, C.~{Charbonnel}, \& R.~F. {Wing}, 92

\bibitem[{Zaritsky {et~al.}(1994)Zaritsky, Kennicutt, \& Huchra}]{Zaritsky1994}
Zaritsky, D., Kennicutt, R.~C., \& Huchra, J.~P. 1994, The Astrophysical
  Journal, 420, 87

\end{thebibliography}

\end{document}